\documentclass[journal]{IEEEtran}

\usepackage{comment}

\usepackage{cite}

\ifCLASSINFOpdf
\usepackage[pdftex]{graphicx}
\usepackage[font=small]{caption}
\usepackage[labelformat=simple,font=scriptsize]{subcaption}

\else
\fi
\usepackage{amsthm}
\usepackage{amssymb}
\usepackage{amsmath}
\usepackage{algorithmicx, algorithm, algpseudocode}

\usepackage{adjustbox}

\begin{document}
		%
		\title{Covert Communication in \\Autoencoder Wireless Systems}
		%
		%
		%
		
		\author{Ali~Mohamamdi~Teshnizi, Majid~Ghaderi, Dennis~Goeckel
			}

	\maketitle
	

\begin{abstract}
Hiding the wireless communication by transmitter Alice to intended receiver Bob from a capable and attentive adversary Willie has been widely studied under the moniker ``covert communications''.  However, when such covert communication is done in the presence of allowable system communications, there has been little study of both hiding the signal and preserving the performance of those allowable communications. Here, by treating Alice, Bob, and Willie as a generator, decoder, and discriminator neural network, we perform joint training in an adversarial setting to yield a covert communication scheme that can be added to any normal autoencoder. The method does not depend on the characteristics of the cover signal or the type of channel and it is developed for both single-user and multi-user systems. Numerical results indicate that we are able to establish a reliable undetectable channel between Alice and Bob, regardless of the cover signal or type of fading, and that the signal causes almost no disturbance to the ongoing normal operation of the system.
\end{abstract}


%
\IEEEpeerreviewmaketitle

	\section{Introduction}
%
%
%
%
\IEEEPARstart{D}{ue} to the broadcast nature of wireless channels, considerable attention has been given to the security and privacy aspects of wireless communications. While the content of messages (i.e., the information transmitted over the channel) can be protected against unauthorized access using cryptography or physical-layer security techniques \cite{zhou2013physical}, there are occasions when hiding the very existence of the communication channel is as vital as securing the communicated messages themselves. Examples of such situations include military operations, cyber espionage, social unrest, or privacy concerns of communication parties. All of the aforementioned use cases have motivated the study of hidden communication channels, which are referred to as ``covert communication'' \cite{lampson1973note, bloch2016covert} in the literature.

The preliminary attempt to obtain covertness started with the study of spread spectrum almost a century ago, with the main purpose of hiding military communications \cite{scholtz1982origins}. The idea was to transmit the signal over a wide frequency band, which would make it harder to locate and identify the original signal amidst the background noise. Many works continued to further examine different aspects of this idea \cite{reynders2016chirp, yan2019low}. However, the fundamental performance limits of such work were unknown until recently when Bash et al. \cite{bash2013limits} established a square root limit on the number of covert bits that can be reliably sent over an additive white Gaussian noise (AWGN) channel. Following this work, there has been a surge of interest in examining covert channels \cite{sobers2017covert,soltani2018covert,sheikholeslami2018multi,cao2018wireless}.

Numerous works have studied the theoretical limits of covert communication over wireless systems in different scenarios \cite{bash2012square, soltani2018covert, sheikholeslami2018multi, li2021fundamental}, but only a few works have focused on the practical implementation of covert communication \cite{dutta2012secret, cao2018wireless, liao2020generative, mohammed2021adversarial}. Many of these works involve an external factor that covert users rely on to build their covert communication, such as hardware impairments \cite{mohammed2021adversarial}, the presence of a cooperative jammer \cite{sobers2017covert}, or the cooperation of a relay node \cite{liao2020generative, kim2022covert}. Additionally, the majority of works make some favorable assumptions for covert users, the accessibility of covert users to cover signals and modulation type \cite{grzesiak2021wireless}, uncertainty in the knowledge of noise power at the detector's receiver \cite{he2017covert}, neglecting the impact of the covert system on normal communication \cite{mohammed2021adversarial}, and limiting the channel model to AWGN \cite{mohammed2021adversarial}. Imposing such restricted assumptions and dependencies eliminates the generality of these covert models and makes it difficult for them to adapt to different system deployments with distinct conditions. Moreover, recent studies show that covert communication that causes noticeable divergence in the statistical properties of signals can be easily detected using analytical and steganalysis methods \cite{bahramali2021robust, huang2020exploiting}. In our work, we leverage the inherent channel noise present in common wireless communication systems. More specifically, our covert communication relies on an established implicit shared secret key between the covert sender and receiver, unbeknownst to the observer. This shared secret key is carefully formed through the joint training of covert sender and receiver as encoder and decoder networks.

There is a large body of work that focuses on different schemes for covert communication based on traditional design approaches. However, wireless systems research has greatly expanded in recent years to consider the potential impact of machine learning (ML) approaches for a variety of problems \cite{wang2017deep}. In fact, various network optimization problems, which were traditionally handled using statistical models, now leverage machine learning techniques \cite{zhu2020toward}. Deep neural networks (DNNs) have helped address several wireless problems, such as signal classification \cite{o2016radio, o2017introduction, wu2020deep, makkuva2021ko}, channel estimation \cite{soltani2019deep}, transmitter identification \cite{roy2019rfal, hanna2019deep}, jamming, and anti-jamming \cite{arjoune2020novel, bahramali2021robust}. A recent study introduced an end-to-end communication model based on deep learning, replacing conventional modular-based designs \cite{o2017introduction}. In this new paradigm, the transmitter and receiver are designed based on DNNs that are jointly trained as the encoder and decoder of an autoencoder network \cite{o2017introduction}. The autoencoder network learns signal characteristics and develops modulation and coding techniques through self-learning, eliminating the need for hand-crafted methods and enabling the system to optimize modulation and encoding strategies. Compared to traditional communication systems, this approach offers greater flexibility and robustness, as the autoencoder can adapt to varying channel conditions and noise levels without manual tuning \cite{zou2021channel}. Despite these benefits, Autoencoder wireless systems, similar to many other deep learning models, are highly susceptible to adversarial attacks \cite{chakraborty2018adversarial}, such as jamming \cite{bahramali2021robust}, spoofing attacks \cite{shi2020generative}, signal misclassification \cite{sadeghi2019physical}. This motivates us to study the vulnerabilities that ML-based wireless communication systems might have against covert communications.

In this work, we introduce a novel deep learning-based covert communication scheme that is free of many of the aforementioned assumptions. It merely relies on the existing channel's noise effect and is independent of any external factor. Our scheme also requires no knowledge of the cover signals or the modulation type. Remarkably, in communication scenarios with a single normal user, our scheme does not even need knowledge about the type of the communication channel. Finally, by training our covert models in an adversarial setting against the observer and minimizing the impact of the covert communication on ongoing normal communication, we ensure that our added covert signals do not cause any conspicuous deviation in the statistical properties of the normal signals. This prevents our covert signals from being easily detected using analytical tools. It is also worthwhile to mention that even though we are proposing our method for autoencoder-based wireless communication systems, there is no limitation on integrating our model into existing conventional wireless communication systems.

The contributions of this work can be expressed as follows:
\begin{itemize}
	\item \textbf{Input and Channel Independent GAN-based Covert Model}: We propose a novel covert communication approach using generative adversarial networks (GANs) that utilizes an input-agnostic generator and discriminator network to represent the covert sender and detector, respectively. These components are trained adversarially in a similar fashion to GANs, allowing our scheme to be independent of cover signals, waveforms, and modulation types used in wireless systems. Moreover, we trained our model on three channel models (AWGN, Rayleigh Fading, and Rician Fading) and conducted experiments to demonstrate its adaptability to different channel conditions and robustness against noise levels. This enables our scheme to be trained and operate on the channel without requiring prior knowledge of the channel model or noise characteristics, providing flexibility in real-world scenarios.
	\item \textbf{Achieving a Controllable Trade-off between Covertness and Performance through the Training Algorithm}: We developed a training procedure that enables us to attain any desired trade-off between the level of covertness and the system's performance (i.e., communication rate) regardless of the number of normal users in the system. This is accomplished by utilizing a regularized loss function for covert users, enabling them to prioritize their objectives based on the specific communication scenario.
	\item \textbf{Comprehensive Experimental Results}: We conducted multiple experiments to demonstrate that our scheme can be integrated into communication systems with both single and multiple normal users under various channel conditions. Notably, in all cases, including single and multi-user scenarios for the three channel models, our (8,1) covert model is demonstrated to have a negligible impact on the normal users' BLER, while establishing a reliable covert communication link and consistently deceiving the detector at various SNRs. Furthermore, our experiments highlighted that there is a degree-of-freedom effect in our scheme, where increasing the number of users affects the performance of the covert and normal communication systems in the fading channels.
\end{itemize}

This paper is organized as follows: Section \ref{s:related} provides an overview of related works on image steganography and covert communication. Section \ref{s:back} gives a brief overview of autoencoder wireless systems. Section \ref{s:model} introduces the system model. The GAN-based covert design is presented in Section \ref{s:gan_covert}. Section \ref{s:eval} presents the experiments and evaluation results. Finally, Section VII concludes the paper and discusses future research directions.

	\section{Related Works}
\label{s:related}
Since the main idea of our work stems from steganography techniques, we first briefly go over the history and current state of this field of research. We then continue this section by reviewing some of the existing approaches to establishing covert communication at the physical layer of wireless networks.

\textbf{Image Steganography}: Deep learning algorithms have greatly advanced various fields, including steganography. Convolutional neural networks (CNNs), originally used in computer vision tasks, have demonstrated remarkable performance in image steganalysis, surpassing traditional statistical methods \cite{tan2014stacked, qian2015deep, xu2016structural}. One of the earliest works of image steganography using deep neural networks is by Baluja \cite{baluja2017hiding}. In this work, Baluja proposes a hiding scheme in which the three networks of preparation, hiding, and reveal sort out the secret encoding and decoding task. The preparation network transforms the hidden message into compressed image features. The hiding network then embeds these features into the cover image, which is processed by the reveal network to extract the secret message. Subsequent research has found that the preparation network can be omitted, simplifying the framework without compromising its effectiveness \cite{zhang2021brief}. The disadvantage of these schemes is that the encoding process is reliant on the cover image. To address this, Zhang et al. \cite{zhang2020udh} propose a new architecture in which the secret message can be encoded independently of the cover image. Besides having more flexibility in hiding the information, this approach has also become an effective method for image watermarking. To manifest robustness against steganalysis practices, researchers started to adopt GAN architectures \cite{goodfellow2014generative}. Volkhonskiy et al. \cite{volkhonskiy2020steganographic} propose one of the first steganography techniques based on GAN networks. The main idea of their work is to use a generative network to produce a new set of cover images that, when carrying the secret message using any of the available steganography techniques, will be less exposed to be detected by a discriminator network (i.e., a steganalysis network). Similarly, Hayes et al. \cite{hayes2017generating} introduce a GAN-based steganography technique with a different objective for the generator network. Instead of generating cover images, the generator learns to embed secret messages into the cover images so that the discriminator cannot distinguish cover images from steganographic images. Although this adversarial scenario was initially introduced for hiding data in images, researchers found it so versatile that it has been applied to other forms of data such as video, audio, and text \cite{martin2023evolving}. This has inspired us to investigate the applicability of such techniques in the wireless communication domain.

\textbf{Covert Communication}: Covert communication has been applied to almost every network layer and protocol, such as IP \cite{cabuk2004ip}, MAC \cite{sheikholeslami2020covert}, and DNS \cite{nussbaum2009robust}. However, covert communication at the physical layer has received little attention until recently. Dutta et al. \cite{dutta2012secret} are one of the first researchers who developed a covert communication technique for the physical layer of wireless communication systems. They leverage communication noise, which can be caused by either the channel or hardware imperfections, as a means to establish the covert channel. In their method, messages are covertly encoded in the constellation error of the normal cover signals. Similarly, Cao et al. \cite{cao2018wireless} further improved this method with the goal of reducing the probability of detection. However, the distortions that these methods cause in the statistical properties of the system were later found not to be difficult to detect using steganalysis methods \cite{huang2020exploiting}, which in turn compromises the covert channel. More recent works have explored the viability of deep neural networks in the covert communication problem. Sankhe et al. \cite{sankhe2019impairment} propose a method called Impairment Shift Keying that produces subtle variations in normal signals in a controlled way such that a CNN model can be trained to classify them as zeros or ones. Although the impact of their covert method on the system's communication error rate is as small as 1\%, authors in \cite{huang2021detection} showed that even tiny modifications to the signals' constellation points can be detected using a CNN model trained on the amplitude and phase characteristics of the error vector magnitude (EVM) and constellation points. Besides, their scheme relies on existing hardware impairments in the system, which is not the case in many deployments. To find an optimal solution for the highest covert rate and minimum probability of detection, Liao et al. \cite{liao2020generative} employed a GAN model that can adaptively adjust the signal power at a relay station for establishing covert communication. This requires the covert users to have access to a cooperative relay node, which is not the case in many communication scenarios. Another example of adversarial training for covert communication can be found in \cite{kim2022covert}. Their setup contains a transmitter communicating with a receiver through a Reconfigurable Intelligent Surface (RIS), and their goal is to keep this communication covert from a prospective eavesdropper. Both the intended receiver and the eavesdropper use CNN classifiers to detect the signals. This scenario raises the same concern as the previous work, which is the necessity of the existence of a relay node in the deployment. Moreover, perturbations added to the signals to deceive the eavesdropper are crafted using the fast gradient method (FGM) \cite{goodfellow2014explaining}, which in \cite{bahramali2021robust} was shown to be easy to counter using existing countermeasure techniques. Our work differs from these two works in two key aspects. First, our proposed method does not rely on any external entities or external factors, such as relay nodes or hardware impairments, which helps with the generality of our model. Second, we have specifically designed our covert model to have as little impact on normal communication as possible, ensuring that the error rate of normal communication does not suspiciously rise. This objective is critical and often overlooked in previous works, which could easily expose any covert communication to the system's observer.

One of the most related works to ours is the covert scheme proposed by Mohammed et al. \cite{mohammed2021adversarial}, where covert communication is formulated as a three-player adversarial game. In their setup, the encoder and decoder networks learn to covertly communicate through noise, while a detector network aims to differentiate covert communication from normal transmissions. While our method shares some ideas with this work, it addresses several critical limitations. First, our model embeds covert signals into the existing channel's noise instead of relying on hardware impairment noise. Second, we optimize our model to preserve the performance of normal communication. Lastly, in contrast to the previous work, which assumes only AWGN channel model, our approach is robust and considers fading channel models like Rayleigh and Rician fading channels, in addition to AWGN.
	\section{Autoencoder Wireless Systems}
\label{s:back}

\begin{figure}[tp!]
	\center
	\includegraphics[width=.7\linewidth]{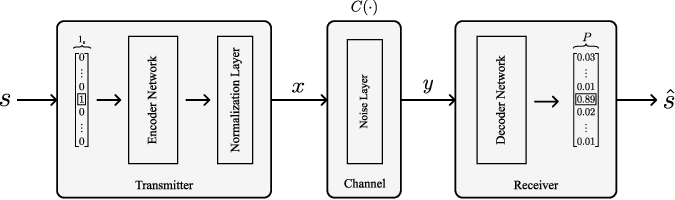}
	\caption{An end-to-end autoencoder-based communication system. The system receives a message \(s\) as input and generates a probability distribution over all possible messages. The most probable message is then selected as the output \(\hat{s}\) \cite{o2017introduction}.}	
	\label{fig:original_autoencoder_architecture}
\end{figure}

Autoencoder-based communication systems are a relatively new development in wireless communications and have numerous advantages, such as their ability to learn from data and adapt to changing conditions in the wireless environment, making them ideal for dynamic wireless environments. They can also be trained to handle noise and errors in the transmitted data, not only for linear, but also for nonlinear channel effects. Autoencoder-based wireless communication systems replace traditional modular components with DNNs, providing an end-to-end learning paradigm. The encoder learns the statistical properties of the wireless channel and modulates the signal, while the decoder reconstructs the message from the received noisy signal. As shown in Fig. \ref{fig:original_autoencoder_architecture}, these systems use autoencoder neural network designs \cite{baldi2012autoencoders} to enhance the performance of wireless communication systems. We describe below how this communication works in both single-user and multi-user systems in greater detail.

\textbf{Single-User Autonencoder Systems}: In this type of communication system, the encoder transforms \(k\) bits of data into a message \(s\) where \(s \in \{1,...,M\}\) and \(M = 2^k\). The encoder then takes this transformed message as an input and generates a signal \(X = E(s) \in \mathbb{R}^{2n}\), which is a real-valued vector. This \(2 \times n\)-dimensional real-valued vector can be treated as an \(n\)-dimensional complex vector, where \(n\) is the number of channel uses required for signal transmission. To account for channel noise, usually additive white Gaussian noise (AWGN), the noise effect is added to the signal vector, denoted as \(N\), which is a \(n\)-dimensional independent and identically distributed (i.i.d.) vector with each element coming from a complex normal distribution with 0 mean and \(\sigma^2\) variance \(N_i \sim \mathcal{CN}(0, \sigma^2)\). A fading coefficient \(H\) is introduced to account for the varying channel conditions. If there is no fading, \(H\) is equal to the identity matrix \(I_n\). However, in the presence of fading, \(H\) is a diagonal matrix with each element following a fading distribution, such as Rayleigh or Rician. Ultimately, the received signal at the receiver, carrying the channel's noise, can be expressed as \(Y = H \cdot X + N\). Once the signals pass through the channel, the decoder receives these distorted signals and applies the transformation \(D: \mathbb{R}^{2n} \rightarrow M \) to output the reconstructed version of the message \(s\), which is  denoted as \(\hat{s} = D(Y)\).

\textbf{Multi-User Autonencoder Systems}: A multi-user autoencoder communication system is an extended version of a single-user system with either multiple transmitters and receivers or multiple transmitters and a central receiver. In this work, we consider the latter case. In this system, each encoder sends a separate message \(s_i\) where \(s_i \in s\), and \(i\) is the index of the transmitter. Each transmitter generates a modulated signal accordingly: \(X_i = E_i(s_i) \in \mathbb{R}^{2n}\). We consider a multiple-access system, where all transmitters send their signals at the same time. Consequently, the signals experience channel interference in addition to the channel effects. When passing through the channel, signals are simultaneously faded and interfere with each other. The resulting vector of signals for each transmitter can be expressed as \(Y_i = \sum_{i=1}^{n_{tx}} H_i \cdot X_i \cdot e^{j\theta_i} + N_i\), where \(H_i\) is the channel coefficient and \(e^{j\theta_i}\) is the phase offset for the \(i^{th}\) transmitter, \(X_i\) is the corresponding encoded signal, and \(n_{tx}\) is the number of transmitters. Finally, the signals are received at the decoder, where it uses its decoding function \(D(\cdot)\) along with the channel matrix \(H\) with the size of \(n_{tx} \times n_{rx}\), where \(n_{rx}\) is the number of antennas, to reconstructed the message \(\hat{s} = D(Y, H)\).
	\section{System Model}
\label{s:model}
\begin{figure*}[tp!]
	\center
	\includegraphics[width=.7\linewidth]{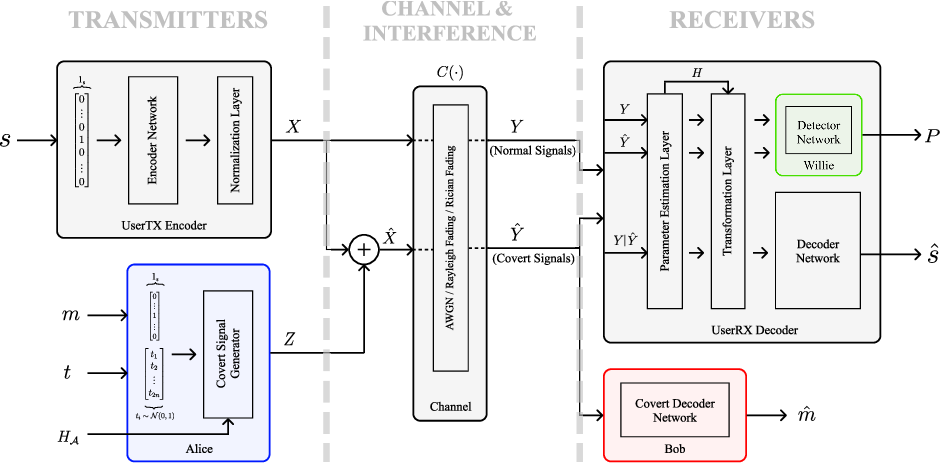}
	\caption{Overall architecture of our system model in the single-user scenario. The UserTX uses an encoder network to encode a binary message $s$ into a vector of signals $X$. Alice generates a covert signal vector $Z$ for a covert message $m$. Willie receives both normal $Y$ and covert $\hat{Y}$ signals during training, but at the time of operation, performs covert detection on either signal based on covert user activity. Similarly, UserRX extracts the normal messages from either $Y$ or $\hat{Y}$. Bob decodes the covert message from $\hat{Y}$. The colored components are accessible to covert users, while the gray components are inaccessible.}	
	\label{fig:system_architecture}
\end{figure*}

Our system comprises normal users communicating through autoencoder wireless systems and covert users aiming to establish hidden communication without arousing suspicion. Communication between normal users can be single-user (i.e., a single transmitter and receiver) or multi-user (i.e., multiple transmitters and a base station receiver). First, we describe the system model in the single-user case, simplifying the system by eliminating interference complexity. This applies to systems handling user interference at higher levels or using multiple access techniques like Orthogonal Frequency Division Multiple Access (OFDMA), Code Division Multiple Access (CDMA), and Time Division Multiple Access (TDMA) \cite{WALRAND2000305}. Next, we continue with a more complex scenario, which is the multi-user case with interfering signal transmissions. Our single-user and multi-user system models are illustrated in Figs. \ref{fig:system_architecture} and \ref{fig:multi_system_architecture}, respectively.

\subsection{Single-User Communication}
\textbf{Transmitter}: In the single-user case, the encoder or the transmitter is referred to as UserTX, and the decoder or the receiver is referred to as UserRX. These two entities together form our autoencoder-based normal wireless communication system. The normal communication process begins with UserTX encoding a binary message to a vector of signals using its encoder network. This vector of signals is then transmitted to UserRX and gets distorted while passing through the channel.

\textbf{Channel}: We consider three channel models of AWGN, Rayleigh fading, and Rician fading. To set the \textit{Signal to Noise Ratio} (SNR) in the AWGN model, we keep the transmitter's average power at unit power and adjust the noise power accordingly. For the fading channels, we consider a flat-frequency block-fading channel model, where each signal vector (codeword) is assumed to be faded independently.

\textbf{Receiver}: A noisy version of the transmitted signal is received at the receiver side, where UserRX extracts the message by decoding the signals. In the case of fading channels, the receiver equalizes the signals before passing them to the decoder network.

\textbf{Equalization}: The channel matrix is estimated by UserRX using a blind channel estimation technique, by feeding the received signals to a preliminary network to predict the fading coefficients. Using the estimated channel matrix, UserRX equalizes the signals prior to feeding them to the decoder network. In Fig. \ref{fig:system_architecture} under the \textit{RECEIVERS} section, you can see the two layers of ``Paramter Estimation Layer'' and ``Transformation Layer'' that are responsible for the signal equalization process.

\subsection{Multi-User Communication}
\textbf{Transmitter}: In the multi-user case, multiple transmitters (UserTXs) send encoded signals to a central receiver (BaseRX). Each UserTX uses its own encoder network to encode a binary message to a vector of signals. The transmitters have no knowledge of each other's messages and share no network parameters. Each UserTX and BaseRX pair forms an autoencoder model, with UserTX serving as the encoder and BaseRX as the decoder. The encoded signals then are simultaneously transmitted over the channel.

\textbf{Channel}: The multi-user case has the same channel models as described in the single-user case, with the exception that signals experience interference due to simultaneous transmission.

\textbf{Receiver}: BaseRX receives the signals from all the transmitters using multiple antennas after they pass through the channel. It decodes the messages in the same way as the UserRX in the single-user case, but it handles the decoding process for all the transmitters' signals.

\textbf{Equalization}: Similar to the single-user case, BaseRX equalizes the signals before passing them to the decoder networks. However, unlike the single-user receiver, we assume that BaseRX has access to the channel matrix. In practice, BaseRX can use a pilot-based channel estimation technique, which results in a much more accurate estimation of the channel matrix compared to the blind channel estimation used in the single-user case.

\begin{figure*}[tp!]
	\center
	\includegraphics[width=.7\linewidth]{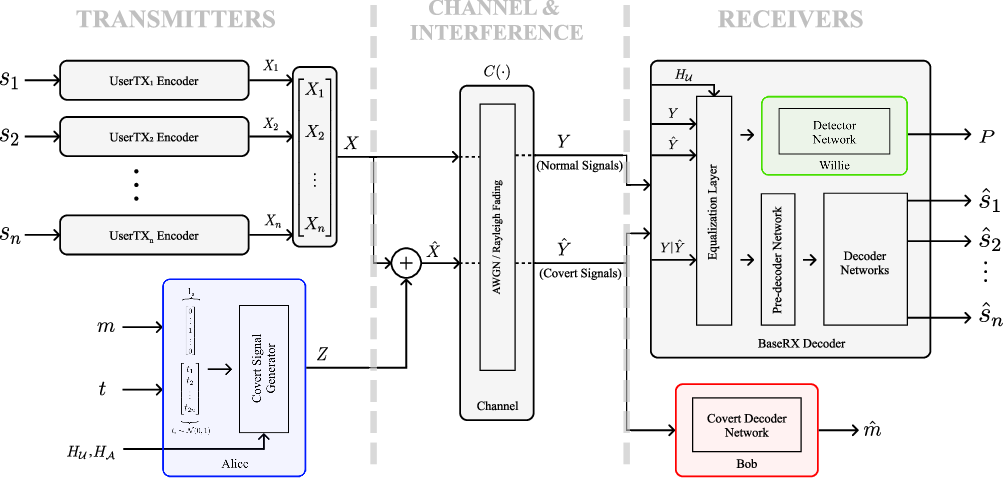}
	\caption{The overall architecture of our system model in the multi-user scenario. Each UserTX separately encodes binary messages $s_i$s into individual signal vectors denoted as $X_i$ ($X$ is a vector of these vectors). Alice generates a covert signal vector $Z$ for a covert message $m$. Willie receives both normal $Y$ and covert $\hat{Y}$ signals during training, but at the time of operation, performs covert detection on either signal based on covert user activity. Similarly, UserRX extracts the normal messages from either $Y$ or $\hat{Y}$. Bob decodes the covert message from $\hat{Y}$. The colored components are accessible to covert users, while the gray components are inaccessible.}
	\label{fig:multi_system_architecture}
\end{figure*}

\subsection{Covert Communication}
There is a covert sender (Alice) who wants to communicate secretly with her intended receiver (Bob). Bob's existence is no secret to the other entities in the system and anyone, including the observer (Willie), can see his transmissions. However, when Willie suspects that a covert sender like Alice is communicating with Bob, he becomes alerted. Therefore, the primary goal of our covert users is to maintain Alice's communications covert.

\textbf{Alice}: The covert communication starts with Alice using her generative model to embed a confidential message into a covert signal vector. You can find an overview of Alice's network in Figs. \ref{fig:system_architecture} and \ref{fig:multi_system_architecture} under the \textit{TRANSMITTERS} sections. As illustrated in the plots, in non-fading channels, Alice merely needs a covert message and a random trigger to produce a covert signal. However, in fading channels, she requires the channel state information of her channel to Bob in the single-user case, and also the channel state information from other normal users to Bob in the multi-user case. This information will be provided by Bob, which we will explain below. After Alice produces the covert signals, she transmits them into the shared channel irrespective of other users' transmissions. This means that her covert signals should be agnostic to the signals of the normal users.

\textbf{Bob}: Unlike the normal receiver or Willie, Bob uses a single antenna at his receiver. He receives the covert signals that have undergone channel effects and interference from normal users' transmissions. Without any equalization operation, he uses his decoder network to extract Alice's message from the covert signals. Additionally, he provides Alice with his and other users' channel information. He does first by sending pilot signals to Alice, and second by measuring channel information using the pilot signals sent from normal transmitters to BaseRX and then broadcasting it to Alice. Bob's existence need not be hidden and him broadcasting this information or pilot signals poses no risk to Alice's secrecy, who is supposed to remain covert.

\textbf{Willie}: Willie listens to all ongoing transmissions on the channel and uses a binary classifier network to determine the likelihood of each signal being covert or normal. Alice must generates covert signals in a way that they do not distort the normal signals, as any distortion can alert Willie to the covert transmission. This requires maintaining the statistical properties of the channel noise and keeping the system's error rate intact. We consider Willie to be an integrated module at the receiver of the normal communication system, i.e., UserRX in the single-user and BaseRX in the multi-user case. This way, he can not only detect incoming covert signals but also measure the communication's error rate.

We represent the roles of Alice, Bob, and Willie with DNNs in a training setup similar to GANs. In this adversarial training setting, each of the three roles is encouraged to maximize its performance until they all reach a state of equilibrium at the end of training.

\section{GAN-Based Covert Design}
\label{s:gan_covert}
For a given binary secret message \(m\), Alice first applies a one-hot encoding technique, and then utilizes her generator model to produce a covert signal \(Z\). With this stochastic generative model, each time a different covert signal is generated for the same message. Alice then sends this covert signal to the shared channel, which is accessible to all entities within the system. To simplify notations, we use \(\hat{X}\) to denote the covert signal before propagation.

\begin{equation}
	\hat{X} = X + Z.
\end{equation}

We consider three channel models: AWGN, Rayleigh fading, and Rician fading. Therefore, there will be three different channel outputs for these three channel models. We use a mapping function \(\mathcal{C}(\cdot)\) to express each of these channels' outputs. Since signals in the multi-user case also experience interference, we express single-user and multi-user channel's outputs separately.

\textit{AWGN Channel Output}: For the AWGN channel model, the signal received at the receiver carries the channel's noise effect \(N \sim \mathcal{N}(0, \sigma^2)\). Consequently, the channel function \(\mathcal{C}(\cdot)\) and the final covert signal \(\hat{Y}\) in the single-user case can be represented as:
\begin{equation}
	\hat{Y} = \mathcal{C}(\hat{X}) = X + Z + N.
\end{equation}
For the multi-user case the covert signal can be denoted as:
\begin{equation}
	\hat{Y} = \mathcal{C}(\hat{X}) = \sum_{i \in U}X_i + Z + N.
\end{equation}
where \(U\) is the set that contains all transmitters.

\textit{Rayleigh and Rician Fading Channel Outputs}: In Rayleigh and Rician fading channels, transmitted signals are multiplied by a complex Gaussian random variable with zero mean and a certain variance, which represents the fading effect caused by multipath propagation. Let \(H_{\mathcal{U}}\) be the complex fading coefficient(s) for the normal signal vector(s), and \(H_{\mathcal{A}}\) be the complex fading coefficient for Alice's signal. The channel function \(\mathcal{C}(\cdot)\) and the resulting covert signal \(\hat{Y}\) in the single-user case are given by:
\begin{equation}
	\hat{Y} = \mathcal{C}(\hat{X}) = (H_{\mathcal{U}} \cdot X) + (H_{\mathcal{A}} \cdot Z) + N.
\end{equation}

In the multi-user case, the received covert signal including the channel interference can be written as:
\begin{equation}
	\hat{Y} = \mathcal{C}(\hat{X}) = \sum_{i \in U}(H_{\mathcal{U}_i} \cdot X_i) + (H_{\mathcal{A}} \cdot Z) + N.
\end{equation}

On the receiving end, Bob receives the covert signal \(\hat{Y}\) and uses his decoder network to reconstruct the covert message \(\hat{m}\).

Willie's network is designed to classify sequences of normal signals \(Y\) and covert signals \(\hat{Y}\). To ensure covert signals remain indistinguishable, Alice generates them to be classified as normal by Willie's network. This requires access to Willie's classifier, which is not feasible in real-world scenarios. Thus, Alice and Bob perform a black-box attack using a substitute binary classifier. Previous studies have shown that adversarial attacks against one model can be effective on another model, even if the two models have different training sets and architectures \cite{papernot2016transferability}. Willie's substitute network will be used to provide feedback to Alice during training, helping her modify the produced covert signals to make them indistinguishable from normally transmitted signals. This ensures that when the model is deployed in a real communication setup, it is highly unlikely that any observer will detect the covert transmissions.

\subsection{General Formulation}
\textbf{Reliability}: The first objective of our covert model is to enable reliable covert communication. In order to achieve this, Bob needs to be able to accurately decode the covert messages sent by Alice. As mentioned earlier, Alice employs a generative model to produce covert signals that correspond to the covert message. Let \(\Theta_{\mathcal{A}}(\cdot)\) be the underlying function of Alice's generative model that takes a random trigger \(t \sim \mathcal{N}(0, 1)\), a covert message \(m\), the channel coefficients from Alice to Bob \(H_{\mathcal{A}}\), and in the multi-user case, the channel matrix \(H_{\mathcal{U}}\) and produces a covert signal \(Z\). The corresponding covert signal can be denoted as \(Z_{m, t} = \Theta_{\mathcal{A}}(m, t, H_{\mathcal{A}}\footnotemark[1], H_{\mathcal{U}}\footnotemark[2])\). Let  \(\Theta_{\mathcal{B}}(\cdot)\) be the underlying function of the decoder network that Bob uses to reconstruct the covert message \(\hat{m}\). Then the reliability of communication between Alice and Bob is achieved using the following loss function:
\begin{equation}
	\begin{aligned} \label{bob_loss}
		\mathcal{L}_{\mathcal{B}} & = \mathbb{E}_{m}[\mathcal{H}(\hat{m}, m)] \\
		& = \mathbb{E}_{m}[\mathcal{H}(\Theta_{\mathcal{B}}(\hat{Y}), m)] \\ 
		& = \mathbb{E}_{m}[\mathcal{H}(\Theta_{\mathcal{B}}(\mathcal{C}(\hat{X})), m)] \\ 
		& = \mathbb{E}_{m}[\mathcal{H}(\Theta_{\mathcal{B}}(\mathcal{C}(\Theta_{\mathcal{A}}(m, t, H_{\mathcal{A}}) + X)), m)].
	\end{aligned}
\end{equation}
For the multi-user case, this equation is written as:
\begin{equation}
	\begin{aligned}
		\mathcal{L}_{\mathcal{B}} & = \mathbb{E}_{m}[\mathcal{H}(\hat{m}, m)] \\
		& = \mathbb{E}_{m}[\mathcal{H}(\Theta_{\mathcal{B}}(\mathcal{C}(\Theta_{\mathcal{A}}(m, t, H_{\mathcal{A}}, H_{\mathcal{U}}) + X)), m)].
	\end{aligned}
\end{equation}
The equation above uses the cross-entropy function \(\mathcal{H}(\cdot)\) to measure the difference between the probability distribution of the reconstructed covert message \(\hat{m}\) and that of the actual covert message \(m\). This equation can be used to optimize the networks of both Alice and Bob by freezing one or the other's network parameters iteratively.

\footnotetext[1]{Parameter only used in fading channels.}
\footnotetext[2]{Parameter only used in multi-user fading channels.}

\textbf{Low Interference}: While (\ref{bob_loss}) ensures communication accuracy, we also need to ensure that the generated perturbations do not negatively impact normal communication between UserTX and UserRX. Otherwise, this could alert Willie to abnormal activity. To address this, we add a constraint that minimizes UserRX's loss function during Alice's training. In a single-user system, we can express this constraint as follows:
\begin{equation}
	\begin{aligned} \label{alice_user_loss}
		\mathcal{L}_{\mathcal{U}} & = \mathbb{E}_{m}[\mathcal{H}(\hat{s}, s)] \\
		& = \mathbb{E}_{m}[\mathcal{H}(\Theta_{\mathcal{U}}(\hat{Y}), s)] \\
		& = \mathbb{E}_{m}[\mathcal{H}(\Theta_{\mathcal{U}}(\mathcal{C}(\hat{X})), s)] \\
		& = \mathbb{E}_{m}[\mathcal{H}(\Theta_{\mathcal{U}}(\mathcal{C}(\Theta_{\mathcal{A}}(m, t, H_{\mathcal{A}}) + X)), s)].
	\end{aligned}
\end{equation}
where \(\Theta_{\mathcal{U}}(\cdot)\) refers to UserRX's decoder network function. It is important to note that during this training, UserRX's decoder network is frozen and only Alice's parameters will be updated.

For the multi-user case, since we have multiple transmitters sending signals, we need to minimize BaseRX's loss function over all individual transmitters' signals. Thus, (\ref{alice_user_loss}) is rewritten as follows:
\begin{equation}
	\begin{aligned} \label{multi_alice_user_loss}
		\mathcal{L}_{\mathcal{U}} & = \sum_{i \in U}\mathbb{E}_{m}[\mathcal{H}(\hat{s}_i, s_i)] \\
		& = \sum_{i \in U}  
		\mathbb{E}_{m}[\mathcal{H}
		(\Theta_{\mathcal{U}}(\mathcal{C}(\Theta_{\mathcal{A}}(m, t, H_{\mathcal{A}}, H_{\mathcal{U}}) + X),  H_{\mathcal{U}}), s_i)].
	\end{aligned}
\end{equation}

\textbf{Undetectability}: In our model, Willie acts as a critic network, similar to what is known as a discriminator in GAN models \cite{goodfellow2014generative}. The so-called real and fake samples in a GANs' discriminator training process here are mapped to normal and covert signals, respectively. For the single-user case, we use the following loss function to optimize Willie's network that classifies covert and normal signals:
\begin{equation}
	\begin{aligned} \label{willie_loss}
		\mathcal{L}_{\mathcal{W}} & = \mathbb{E}_{m}[\mathcal{H}(\hat{Y}, Y)] \\
		& = \mathbb{E}_{m}[\mathcal{H}(\Theta_{\mathcal{W}}(\mathcal{C}(\hat{X})), \Theta_{\mathcal{W}}(\mathcal{C}(X))] \\
		& = \mathbb{E}_{m}[\mathcal{H}(\Theta_{\mathcal{W}}(\mathcal{C}(\Theta_{\mathcal{A}}(m, t, H_{\mathcal{A}}) + X)), \Theta_{\mathcal{W}}(\mathcal{C}(X))].
	\end{aligned}
\end{equation}
where \(\Theta_{\mathcal{W}}(\cdot)\) is the detector function, and \(\mathcal{H}(\cdot)\) is the binary cross entropy between the predicted outputs for the covert signals \(\hat{Y}\) or normal signals \(Y\) and their actual outputs. For the multi-user case, we need to optimize Willie's network over all the transmitters' outputs.

\begin{equation}
	\begin{aligned} \label{multi_willie_loss}
		\mathcal{L}_{\mathcal{W}} & = 
		\sum_{i \in U} \mathbb{E}_{m}[\mathcal{H}(\hat{Y}, Y)] \\
		= \sum_{i \in U} & 
		\mathbb{E}_{m}[\mathcal{H}(\Theta_{\mathcal{W}}(\mathcal{C}(\Theta_{\mathcal{A}}(m, t, H_{\mathcal{A}}, H_{\mathcal{U}}) + X)), \Theta_{\mathcal{W}}(\mathcal{C}(X))].
	\end{aligned}
\end{equation}

\begin{algorithm}[tp!]
	\caption{Covert Model Training}\label{alg:train}
	\small
	\begin{algorithmic}
		\State $X \gets$ normal signals data
		\State $S, M \gets$ normal and covert messages sets
		\State $\Theta_{\mathcal{A}}, \Theta_{\mathcal{B}}, \Theta_{\mathcal{W}} \gets$ Alice, Bob, and Willie networks
		\State $\Theta_{\mathcal{U}} \gets$ UserRX decoder network
		\State $\mathcal{H} \gets$ cross entropy 
		\State $\mathcal{C} \gets$ channel mapping function
		\For{epoch $ep \in \{1 \ldots n_{epochs}$\}}
		\State $t \sim \mathcal{N}(0, 1)$
		\State $\mathcal{L}_{\mathcal{W}} = \mathcal{H}(\Theta_{\mathcal{W}}(\mathcal{C}(\Theta_{\mathcal{A}}(M, t, H_{\mathcal{A}}, H_{\mathcal{U}}) + X)), \Theta_{\mathcal{W}}(\mathcal{C}(X))$
		\State Update $\Theta_{\mathcal{W}}$ to minimize $\mathcal{L}_{\mathcal{W}}$
		\State $\mathcal{L}_{\mathcal{B}} = \mathcal{H}(\Theta_{\mathcal{B}}(\mathcal{C}(\Theta_{\mathcal{A}}(M, t, H_{\mathcal{A}}, H_{\mathcal{U}}) + X)), M)$
		\State Update $\Theta_{\mathcal{B}}$ to minimize $\mathcal{L}_{\mathcal{B}}$
		\State $\mathcal{L}_{\mathcal{U}} \gets \mathcal{H}(\Theta_{\mathcal{U}}(\mathcal{C}(\Theta_{\mathcal{A}}(M, t, H_{\mathcal{A}}, H_{\mathcal{U}}) + X)), S)$
		\State
		$\mathcal{L}_{\mathcal{A}} = \lambda_{\mathcal{B}} \mathcal{L}_{\mathcal{B}} + \lambda_{\mathcal{U}} \mathcal{L}_{\mathcal{U}} - \lambda_{\mathcal{W}} \mathcal{L}_{\mathcal{W}}$
		\State Update $\Theta_{\mathcal{A}}$ to minimize $\mathcal{L}_{\mathcal{A}}$
		\EndFor
	\end{algorithmic}
\end{algorithm}

This white-box adversarial training against Alice's network allows Willie's network to distinguish between covert and normal signals when adequately trained. On the other hand, we also ensure that the covert signals generated by Alice maintain the statistical properties of normal signals on the channel to avoid detection. To achieve this undetectability property, we use Willie's network to act as a discriminator network in Alice's optimization function. Alice's training against this network aims to maximize Willie's uncertainty in his predictions, allowing Alice to form covert communication that resembles channel noise while being intelligible to Bob. Overall, Alice's loss function can be expressed as a weighted sum of these three objectives:
\begin{equation}
	\begin{array}{l} \label{alice_loss}
		\mathcal{L}_{\mathcal{A}} = \lambda_{\mathcal{B}} \mathcal{L}_{\mathcal{B}} + \lambda_{\mathcal{U}} \mathcal{L}_{\mathcal{U}} - \lambda_{\mathcal{W}} \mathcal{L}_{\mathcal{W}}.
	\end{array}
\end{equation}
where \(\lambda_{\mathcal{B}}\), \(\lambda_{\mathcal{U}}\), and \(\lambda_{\mathcal{W}}\) are hyperparameters that determine the relative importance of the different objectives in Alice's loss function. The algorithmic steps involved in training our covert models are summarized in Algorithm \ref{alg:train}.

\subsection{Neural Network Architecture}
\textbf{User's Autoencoder Network}: The autoencoder model takes a binary message \(s\) of size \(k\) bits and produces a reconstructed version \(\hat{s}\). The encoder maps the message to a \(2 \times n\) signal vector, where \(n\) is the number of channel uses, using one-hot encoding. In the multi-user case, the resulting vector has a size of \(n_{tx} \times (2 \times n)\). The signals pass through the channel, where the channel mapping function applies its effects. In single-user systems with channel fading, the receiver uses parameter estimation and transformation to equalize signals. A simple transformation function divides signals by the estimated channel fading coefficients. More complex transformations can be used, as described in \cite{o2017introduction}; however optimizing the autoencoder model's performance is beyond the scope of this article. In multi-user systems, the decoder receives channel coefficients, allowing BaseRX to use the zero-forcing technique \cite{garg2010wireless} for signal equalization. In single-user systems, UserRX reconstructs the message using its decoder network. In the multi-user case, BaseRX decodes signals from all transmitters simultaneously by passing them through a pre-decoder network and using separate decoders at the final layers.

\textbf{Alice's Network}: Alice first transforms a covert message \(m\) to its corresponding one-hot encoding representation, where each message belongs to a unique class. She then uses a random trigger \(t\) to randomize the process by which the covert noise signal \(Z\) is produced, along with the channel coefficients \(H_{\mathcal{A}}\) and \(H_{\mathcal{U}}\). For Alice's generator model, we use multiple dense layers with ReLU and Tanh activation functions. The first layer acts as an embedding layer, enlarging the input's domain space. The subsequent fully connected layers extract useful features and perform the encoding process. Finally, the last layer performs a dimension transformation, ensuring that the generated covert signal \(Z\) complies with the dimension of the normal signal \(X\) on the channel.

\textbf{Bob's Network}: Bob receives the covert signal \(\hat{Y}\) that has been affected by the channel, and he feeds it through his decoder network to extract the secret message. Bob's network is more sophisticated than Alice's, as decoding such a distorted signal is a much more complex task. The received message first goes through a wide dense layer with a Tanh activation function, which increases the input's feature space. The data then passes through multiple 1-Dimensional Convolutional (1D Conv) layers, which learn the coding that Alice has developed to encode the covert messages. We have found that using 1D Conv layers helps Bob and Alice achieve better consistency in the accuracy of their communication, especially when the channel model is more complicated (i.e., when there is also fading in the channel). The rest of Bob's decoder network consists of two dense layers that remap the learned feature space to the covert message domain space. As with UserRX's and BaseRX's decoder networks, Bob eventually predicts the covert message by performing a classification on the received signal.

\textbf{Willie's Network}: Willie's task is to distinguish between the normal signal  \(Y\) and the covert  signal \(\hat{Y}\). To achieve this, he uses a neural network with the same architecture as Bob's, except for the last layer, which has a Sigmoid activation function instead of Softmax. The network takes an input signal, either normal or covert, and produces a confidence probability \(P\) indicating the likelihood of the signal being normal. Using the same network architecture for both Bob and Willie ensures a fair competition between them, as they have the same capacity for learning.

	\section{Experiments and Evaluation}
\label{s:eval}
Our experiments consist of two sections. The first section focuses on evaluating the performance of our trained autoencoder wireless networks. These experiments do not involve covert communication and aim to fine-tune the parameters of the autoencoder models for accurate performance in fading environments, in addition to AWGN channels. The second section analyzes the performance of our implemented covert models in single-user and multi-user scenarios, assessing their robustness to various channel models, data rates, and user counts.

\begin{figure*}[t]
	\centering
	\begin{subfigure}{0.32\textwidth}
		\includegraphics[width=\linewidth]{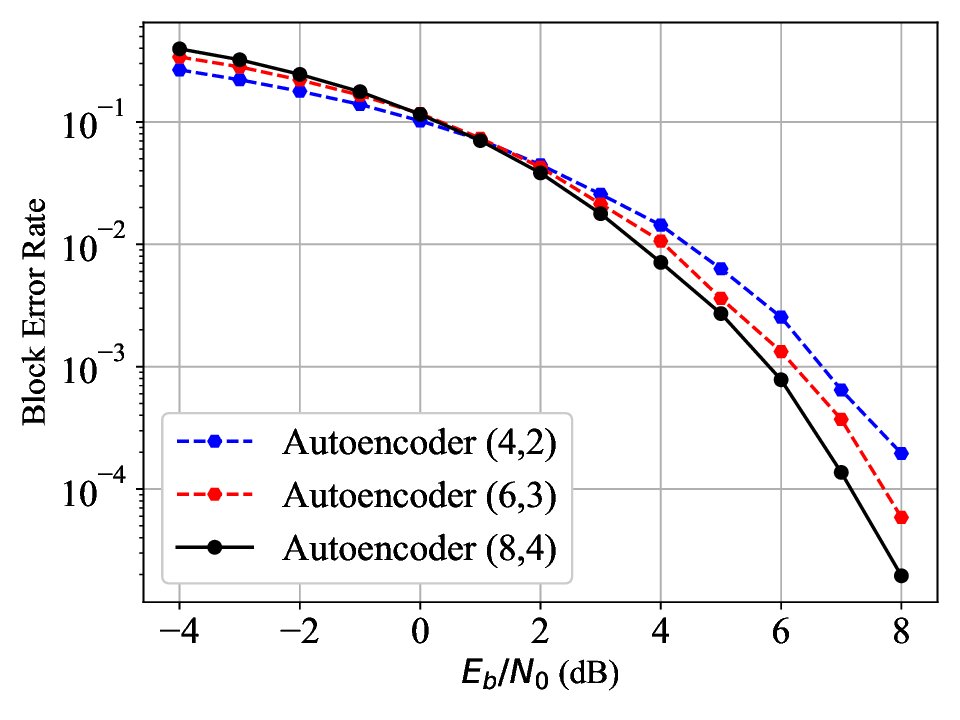}
		\caption{AWGN channel}
	\end{subfigure}
	\begin{subfigure}{0.32\textwidth}
		\includegraphics[width=\linewidth]{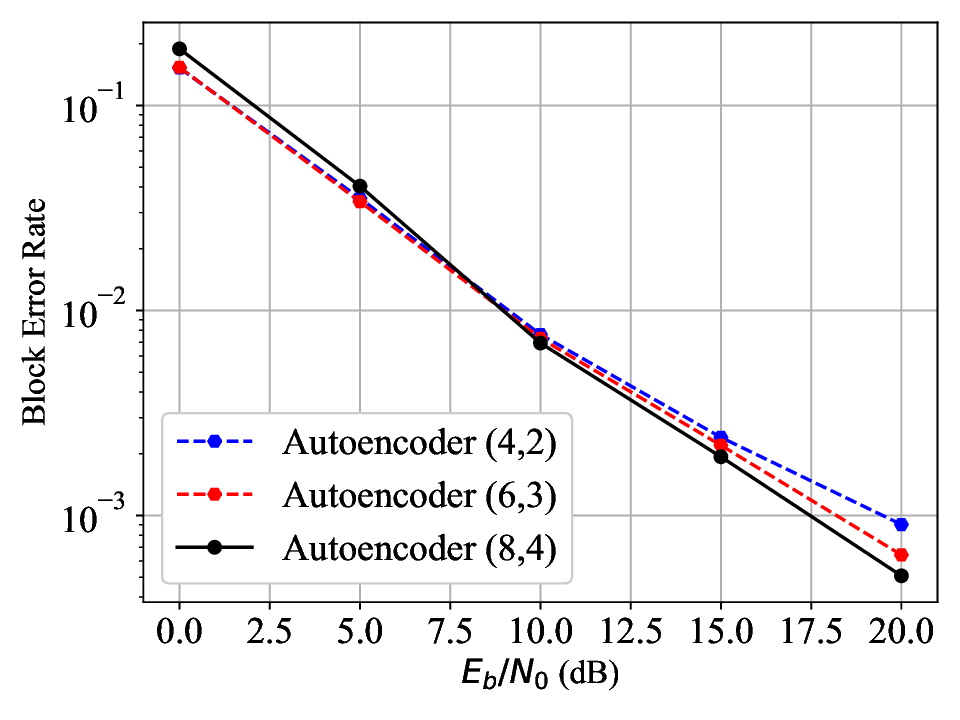}
		\caption{Rician fading channel}	
	\end{subfigure}
	\begin{subfigure}{0.32\textwidth}
		\includegraphics[width=\linewidth]{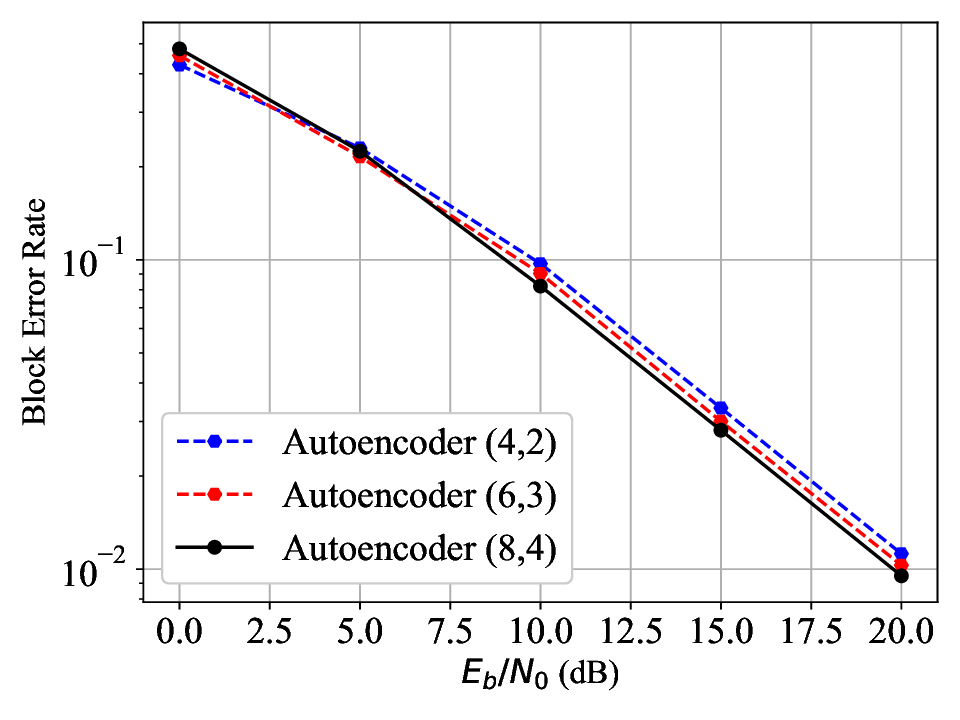}
		\caption{Rayleigh fading channel}	
	\end{subfigure}
	\caption{Autoencoders' performance in terms of BLER over a range of SNR values is evaluated in our single-user case. The models are trained over AWGN, Rayleigh, and Rician fading channels for a set of parameters that have the same data rate.}
	\label{fig:autoencoder_bler}
\end{figure*}

\subsection{Base Single-User Autoencoder's Performance}
\textbf{Methodology}: We implemented an autoencoder communication network for normal communication between UserRX and UserTX. An \(Autoencoder (n, k)\) is a neural network communication model that sends \(k\) bits of data in \(n\) channel uses. To make our results comparable with \cite{o2017introduction}, we chose our default parameters to be 8 and 4 for the number of channel uses and binary message size, respectively. However, we also evaluated our models for two other sets of parameters with the same data rate but different numbers of channel uses. This allowed us to examine how increasing the number of channel uses, or signal dimensionality, would affect communication performance. To train our autoencoder model, we generated two datasets for training and testing by randomly generating binary messages \(s\) of size \(k\). Specifically, we used 8192 random binary messages in the training set and 51200 random binary messages in the test set. We created a much larger dataset for testing to ensure that each signal \(X\) undergoes various channel distortions, providing a more accurate evaluation of the model's performance. We set the learning rate to 0.001 and optimized the model using the Adam optimizer \cite{kingma2014adam}. We used a batch size of 1024 and trained the model for 100 epochs. For the channel configuration, \textbf{we fixed the SNR value during training but evaluated the model's performance over a range of SNRs}. The SNR value for the AWGN channel was set to 4dB, while the values for the Rayleigh and Rician fading channels were 16dB. We chose these SNR values experimentally by training the models on different SNR values and identifying the value on which the model performed best.

\textbf{Results}: Fig. \ref{fig:autoencoder_bler} shows the block error rate (BLER) performance of our trained autoencoder communication models for various sets of parameters across a range of SNR values. The models were trained individually on AWGN, Rayleigh, and Rician fading channels and tested on the same channel they were trained on.
The plot reveals that despite having the same data and coding rate, increasing the signal dimension slightly enhances the performance of the autoencoder models. This phenomenon was first identified in \cite{o2017introduction}, which demonstrated that autoencoders trained over an AWGN channel can achieve a coding gain by learning a joint coding and modulation scheme. Our results support this finding and suggest that this behavior holds true for autoencoders trained on other channel models as well. However, it should be noted that a comprehensive study of the performance of autoencoder wireless systems for multiple channel types and parameters (n, k) goes beyond the scope of this work and is not the primary focus of this research.

\subsection{Base Multi-User Autoencoders' Performance}
\textbf{Methodology}: In the multi-user case, we have chosen the number of channel uses and the binary message size to be 8 and 4, respectively, as these are our default parameters. There are two reasons for selecting these parameters in this way. First, it allows us to compare our results with those obtained in \cite{o2017introduction}. Second, by having each user communicate at half the rate of BPSK, the results of our 2-user system are roughly comparable to a single-user system at the BPSK rate, while the 4-user system is comparable to a single-user system at the QPSK rate. To generate training and testing sets, we followed the same procedure outlined in the previous section. The remaining parameters, such as learning rate, number of epochs, batch size, and optimization algorithm, are kept the same as in the single-user system. For the channel configuration, we have chosen SNR values of 8dB for the AWGN channel, 16dB for the Rayleigh channel, and 14dB for the Rician channel during training. However, we evaluate our models over a range of SNR values.

\begin{figure*}[tp!]
	\centering
	\begin{subfigure}{0.32\linewidth}
		\includegraphics[width=\linewidth]{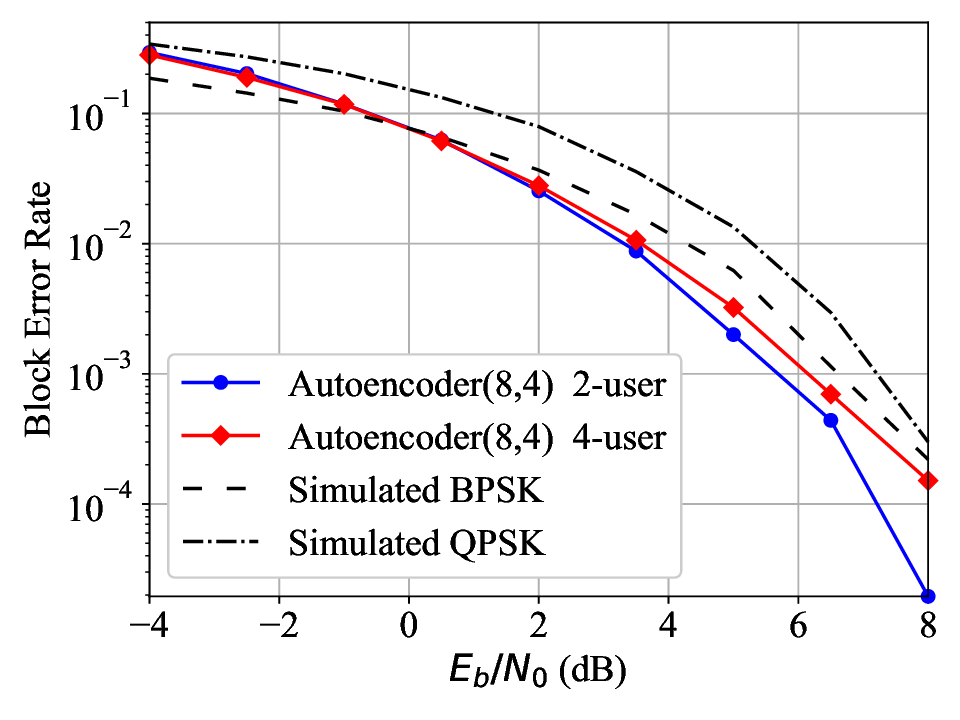}
		\caption{AWGN channel}
	\end{subfigure}
	\begin{subfigure}{0.32\linewidth}
		\includegraphics[width=\linewidth]{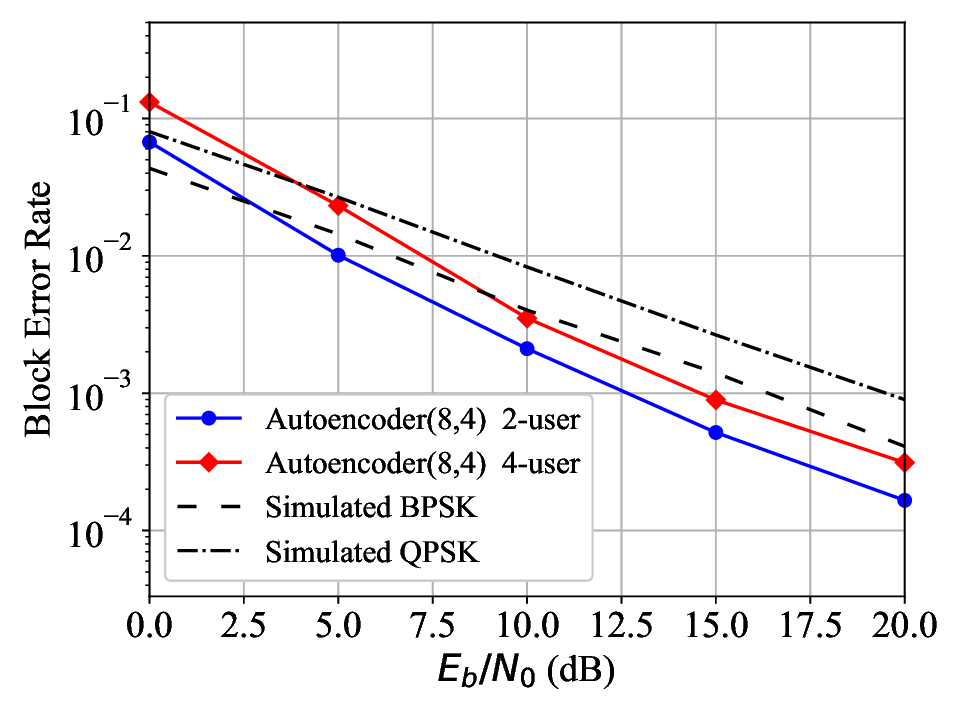}
		\caption{Rician fading channel}	
	\end{subfigure}
	\begin{subfigure}{0.32\linewidth}
		\includegraphics[width=\linewidth]{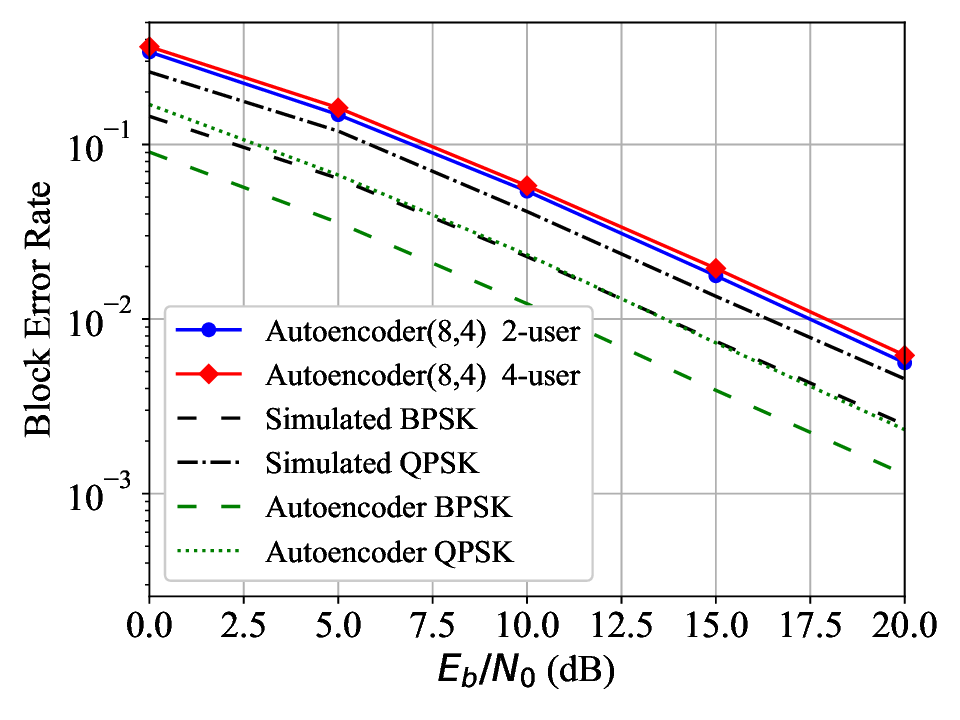}
		\caption{Rayleigh fading channel}	
	\end{subfigure}
	\caption{The block error rates (BLERs) of our trained autoencoders compared with simulated results for different numbers of users over a range of SNR values in our multi-user case.}
	\label{fig:multi_autoencoder_bler}
\end{figure*}

\textbf{Results}: The performance of our trained autoencoder-based communication models in terms of block error rate (BLER) for a range of SNR values for different numbers of users is shown in Fig. \ref{fig:multi_autoencoder_bler}. In all charts, the 2-user and 4-user performances are depicted with blue and red colors, respectively, and the results are compared with simulated traditional BPSK and QPSK systems with hard decision decoding.
The results indicate that multi-user autoencoder models can achieve almost similar performance to their counterparts in the single-user cases when compared data rate-wise. However, we have also observed that while the AWGN and Rician autoencoders outperform their peers, the Rayleigh fading autoencoders do not. We attribute this to the more complex equalization task that the receiver in multi-user cases needs to undertake. This becomes more evident when we compare the results of our trained single-user autoencoders with BPSK and QPSK data rates, which were able to outperform all other results.

\subsection{Covert Model Performance Evaluation}
We evaluated the performance of our covert communication models on three different channel models: AWGN, Rayleigh fading, and Rician fading. We used the same training procedure for all settings, but the network architecture of our covert and autoencoder models in the multi-user case differed slightly from that in the single-user setting. Table \ref{table:covert_models_structure} shows these differences.

\begin{algorithm}[tp!]
	\caption{Optimal SNR Range Search}\label{alg:snr_search}
	\small
	\begin{algorithmic}
		\State $acc_{\mathcal{A, B, W}} \gets$ Alice, Bob, and Willie final training accuracies
		\State $p, c \gets$ Previous and current average training accuracies
		\State $snr_{L, U} \gets$ Optimal lower and upper bounds of the SNR range
		\State $t \gets L$ Tracking the SNR bound that is expanding
		\While{true}
		\State $acc_{\mathcal{A}}, acc_{\mathcal{B}}, acc_{\mathcal{W}} \gets Train(snr_{L}, snr_{U})$
		\State $c \gets Avg(acc_{\mathcal{A}}, acc_{\mathcal{B}}, acc_{\mathcal{W}})$
		\If {$c > p$}
		\State $p \gets c$
		\State $snr_{t} \gets snr_{t} \pm 1$
		\Else
		\If {$t$ is equal $L$}
		\State $t \gets U$
		\Else
		\State \Return $snr_{L, U}$
		\EndIf
		\EndIf
		\EndWhile
	\end{algorithmic}
\end{algorithm}

\textbf{Datasets and Hyperparameters}: Since each covert message \(m\) has to be paired with a normal message \(s\), we created the covert model's training and testing sets to have the same number of samples as the autoencoder's. All models were trained for 5000 epochs using the Adam optimizer in an adversarial training setting. We adjusted the importance of each of Alice's objectives by setting \(\lambda_{\mathcal{W}} = 2 \lambda_{\mathcal{B}} = 4 \lambda_{\mathcal{U}}\) for the single-user case, and \(\lambda_{\mathcal{W}} = 3 \lambda_{\mathcal{B}} = 6 \lambda_{\mathcal{U}}\) for the multi-user case in (\ref{alice_loss}). We arrived at these numbers by running a grid search on these parameters. However, our solution is not limited to these parameters, and one can use a different set of parameters to emphasize one specific objective more than the others. In both the single-user and multi-user cases, we started the training with a learning rate of 0.001 for the first 2500 epochs and then made the learning rate ten times smaller for the remaining 2500 epochs. In each epoch, we first updated the parameters of Willie's network using (\ref{willie_loss}), then trained Alice's network for one step using (\ref{alice_loss}), and finally optimized Bob's network based on (\ref{bob_loss}).

We trained our autoencoder network on a fixed SNR value, but observed improved performance in our covert scheme when training on a range of SNR values. To achieve this, we randomly varied the SNR value within a predetermined range after each training epoch. This approach helped maintain accuracy in normal communication for Alice and improved covert message decoding for Bob, particularly in lower SNR conditions. We started with the training SNR used for the autoencoder and gradually expanded the range from both ends until no further improvement was observed. Algorithm \ref{alg:snr_search} outlines this process. In the single-user case, we settled on the ranges of -2dB to 8dB for AWGN, and 10dB to 30dB for both Rayleigh and Rician fading channels. In the multi-user system, the optimal ranges were found to be 0dB to 10dB for AWGN, 0dB to 20dB for Rician, and 10dB to 30dB for Rayleigh channels.

\begin{figure}
	\centering
	\begin{subfigure}{0.49\columnwidth}
 		\includegraphics[width=\linewidth]{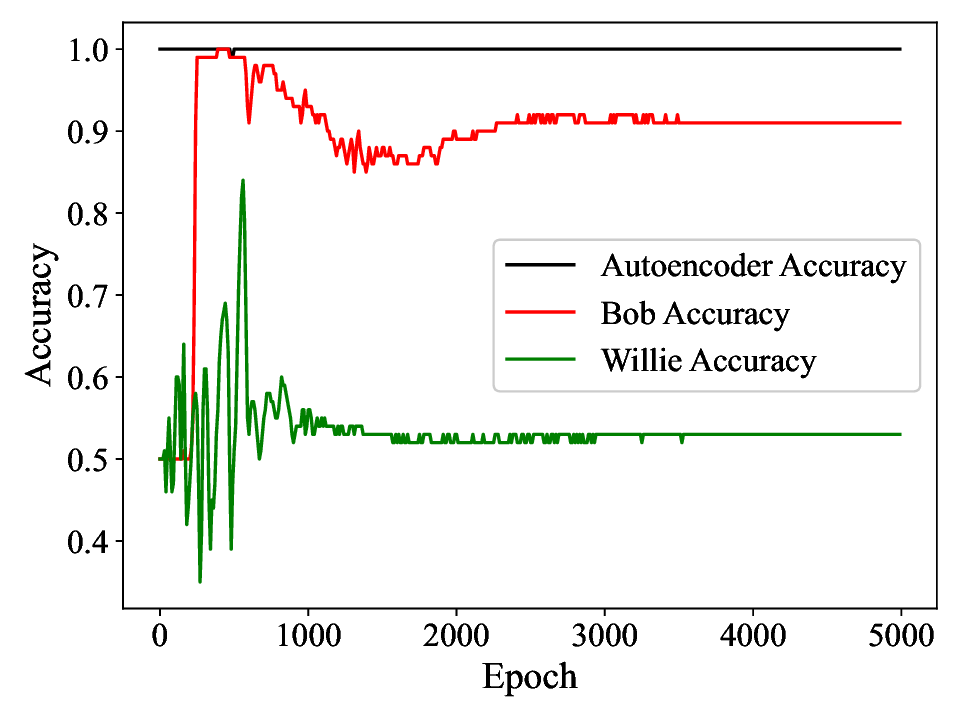}
		\caption{Single-user case}	
	\end{subfigure}
	\begin{subfigure}{0.49\columnwidth}
		\includegraphics[width=\linewidth]{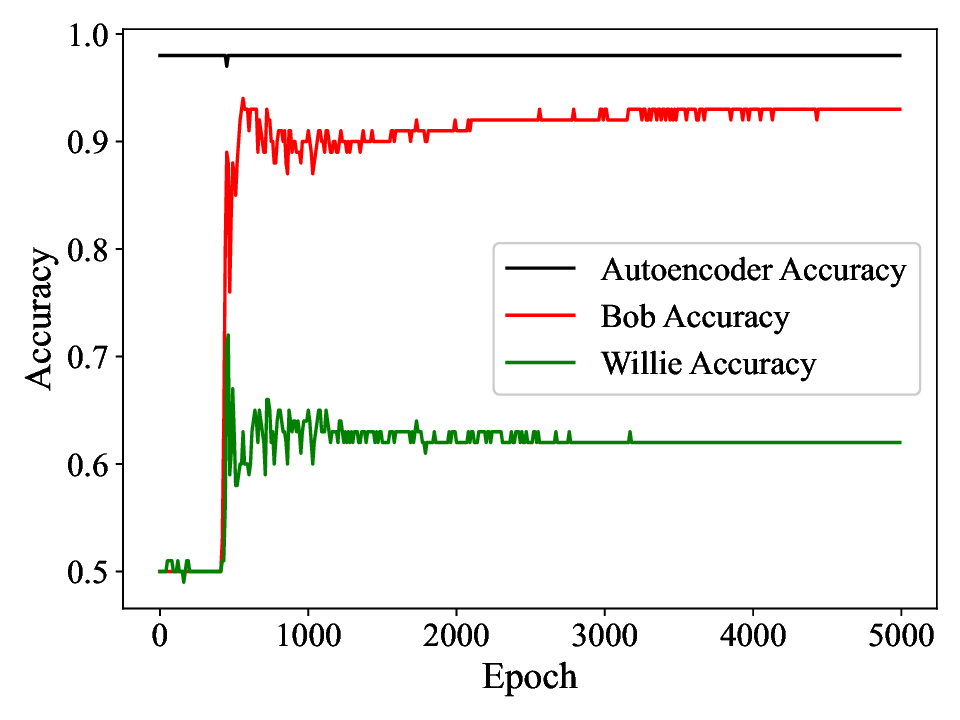}
		\caption{Multi-user case}	
	\end{subfigure}
	\caption{Evaluation results of our covert and autoencoder models during the training process show the system reaches a stable point after successful training.}
	\label{fig:traning_progress}
\end{figure}

\textbf{Training Procedure}: Fig. \ref{fig:traning_progress} shows the progress of each covert actor's accuracy on the test set during the training process for both single-user and multi-user cases. As the training progresses, Bob gradually learns to decode covert messages \(m\) and establishes reliable communication with Alice. After a few epochs as the covert communication begins to take form and stabilize, the signals start to deviate from their original distribution, which helps Willie to better detect covert signals. When Willie's accuracy increases, the term \(\mathcal{L}_{\mathcal{W}}\) dominates the other two objectives of Alice's loss function in (\ref{alice_loss}). As a result, Alice gradually sacrifices accuracy for undetectability. Soon after, the training process reaches a stable point where neither of the covert models sees any significant improvement in accuracy as the training progresses. At the end of the training, the Users' accuracy remains intact, Bob achieves reliable covert communication accuracy, and Willie stabilizes at around 50$\sim$60\% accuracy, which, for a binary classifier, is very close to random guessing accuracy.

\textbf{Single-User Experiments}: 
We started our experiments with the single-user case. First, we evaluated our covert models by sending 1 bit of covert data over 8 channel uses and then gradually increased the number of covert bits to see how increasing the covert data rate affected each component of our covert scheme. We used the notations \(Alice (n,k)\), \(Bob (n,k)\), and \(Willlie (n,k)\) to differentiate models with different covert data rates, and their interpretation was the same as that of the autoencoder model.

\begin{figure*}
\begin{minipage}{\textwidth}
	\begin{subfigure}[b]{0.32\textwidth}
		\includegraphics[width=\linewidth]{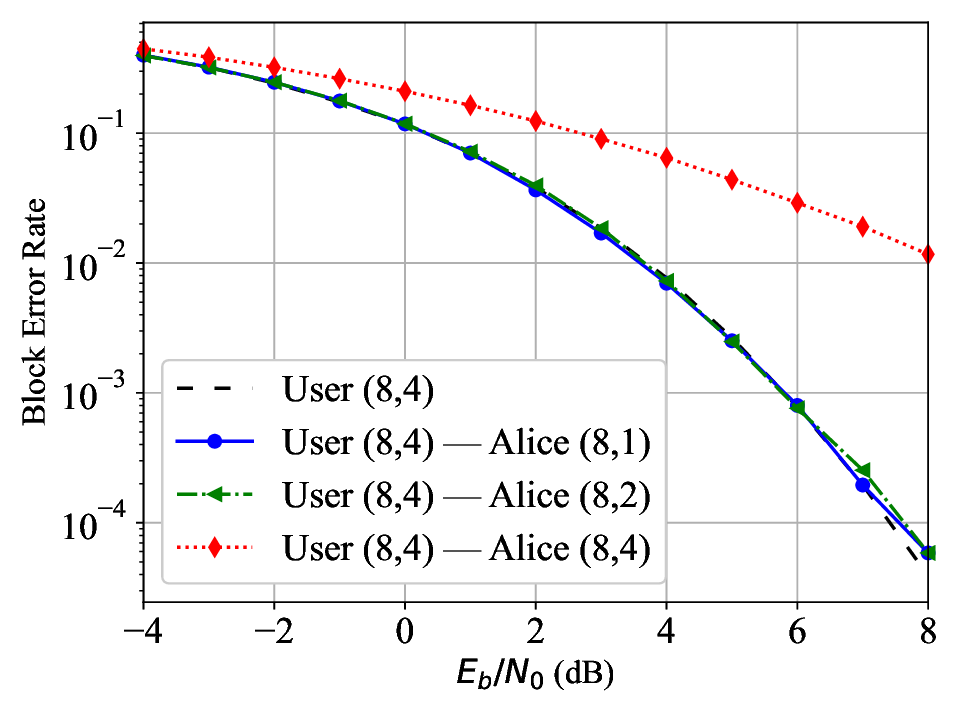}
		\caption{User's BLER}
		\label{fig:awgn_resutls_ae}
	\end{subfigure}\hfill
	\begin{subfigure}[b]{0.32\textwidth}
		\includegraphics[width=\linewidth]{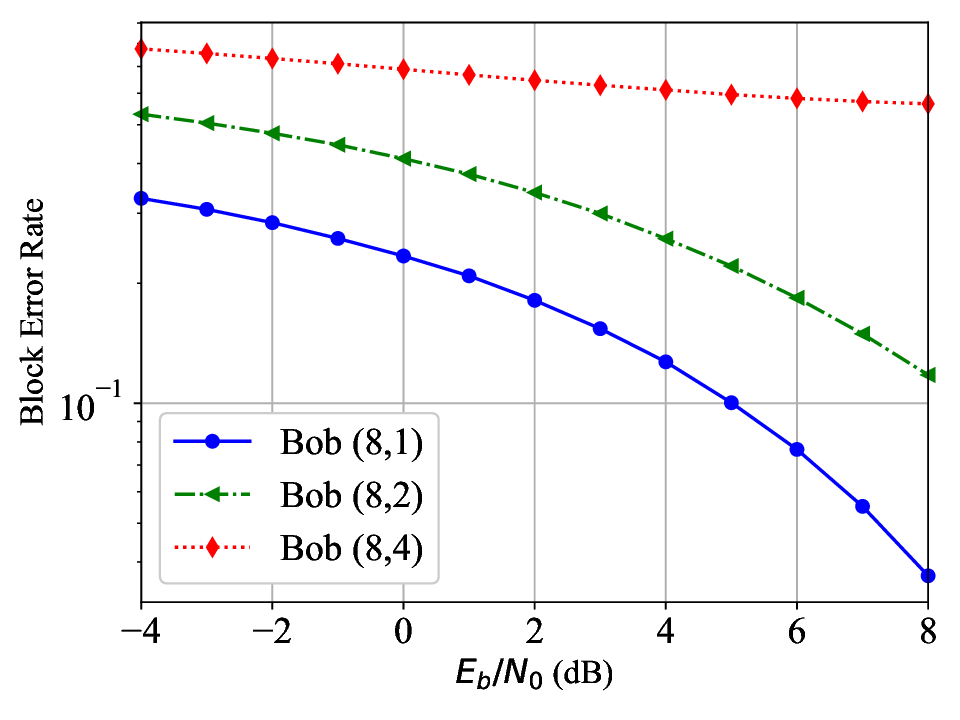}
		\caption{Bob's BLER}	
		\label{fig:awgn_resutls_bob}
	\end{subfigure}\hfill
	\begin{subfigure}[b]{0.32\textwidth}
		\includegraphics[width=\linewidth]{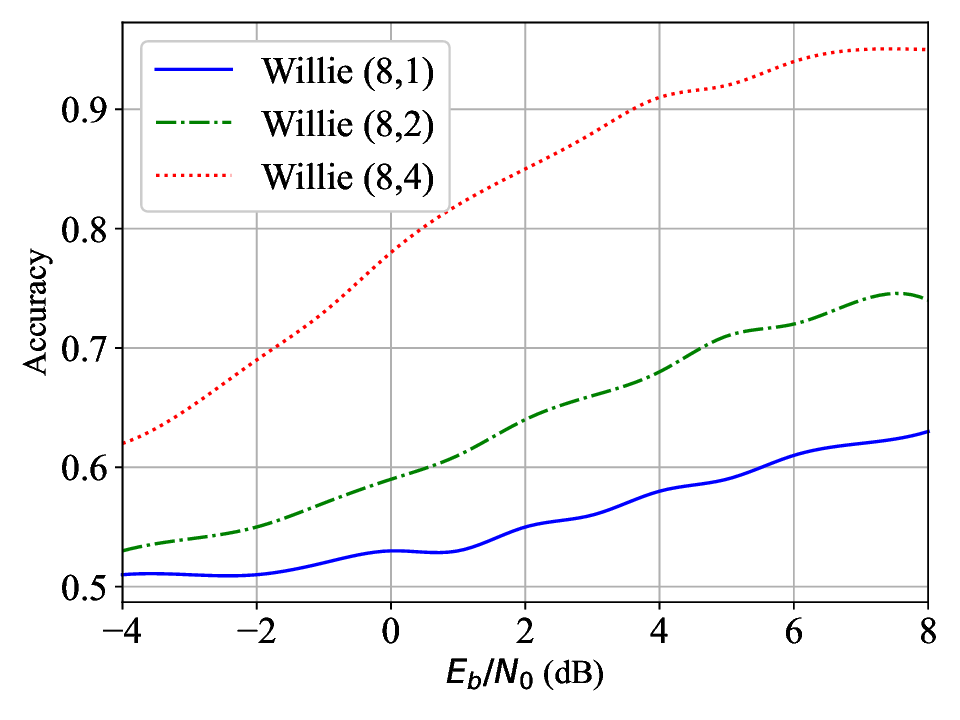}
		\caption{Willie's accuracy}	
		\label{fig:awgn_resutls_willie}
	\end{subfigure}
	\caption{Single-use model performance over AWGN channel for different covert data rates on a range of SNRs.}
	\label{fig:awgn_results}
\end{minipage}
\begin{minipage}{\textwidth}
	\begin{subfigure}[b]{0.32\textwidth}
		\includegraphics[width=\linewidth]{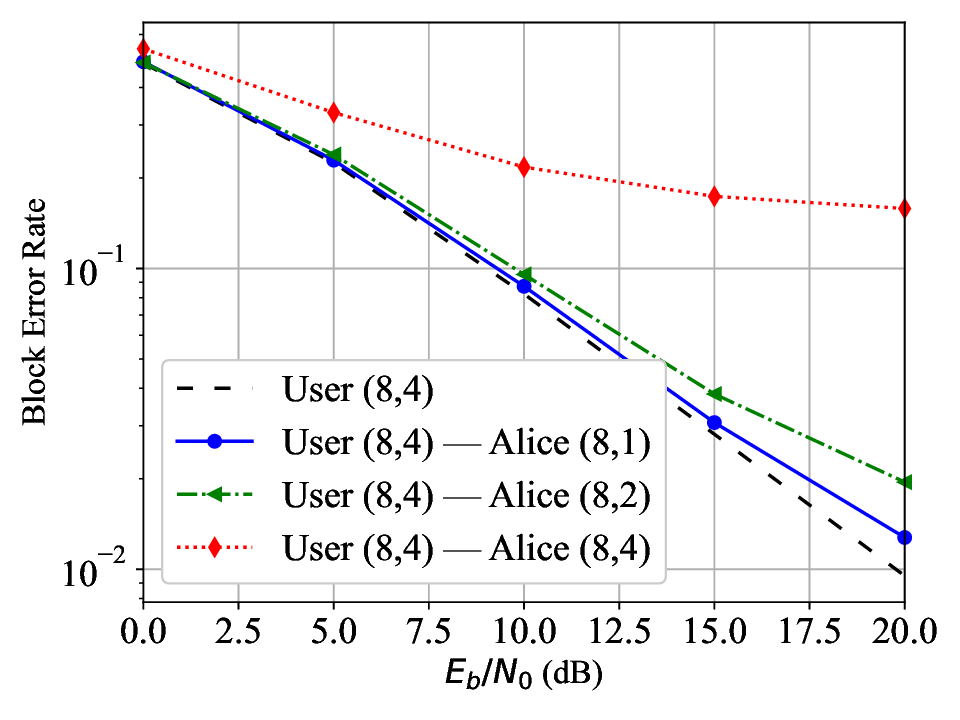}
		\caption{User's BLER}
		\label{fig:rayleigh_resutls_ae}
	\end{subfigure}\hfill
	\begin{subfigure}[b]{0.32\textwidth}
		\includegraphics[width=\linewidth]{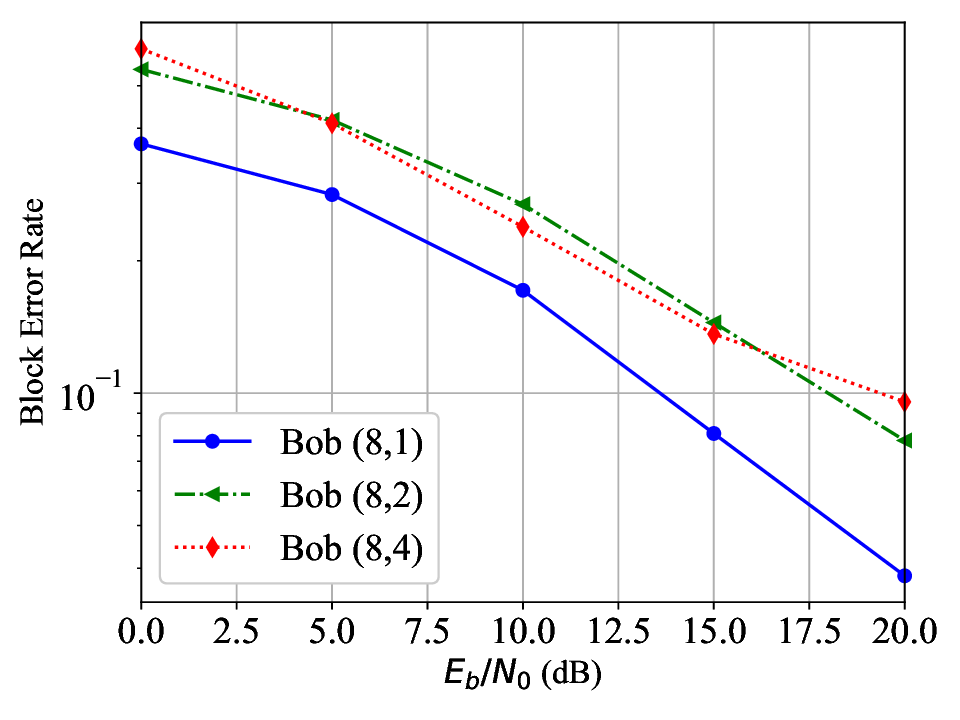}
		\caption{Bob's BLER}
		\label{fig:rayleigh_resutls_bob}	
	\end{subfigure}\hfill
	\begin{subfigure}[b]{0.32\textwidth}
		\includegraphics[width=\linewidth]{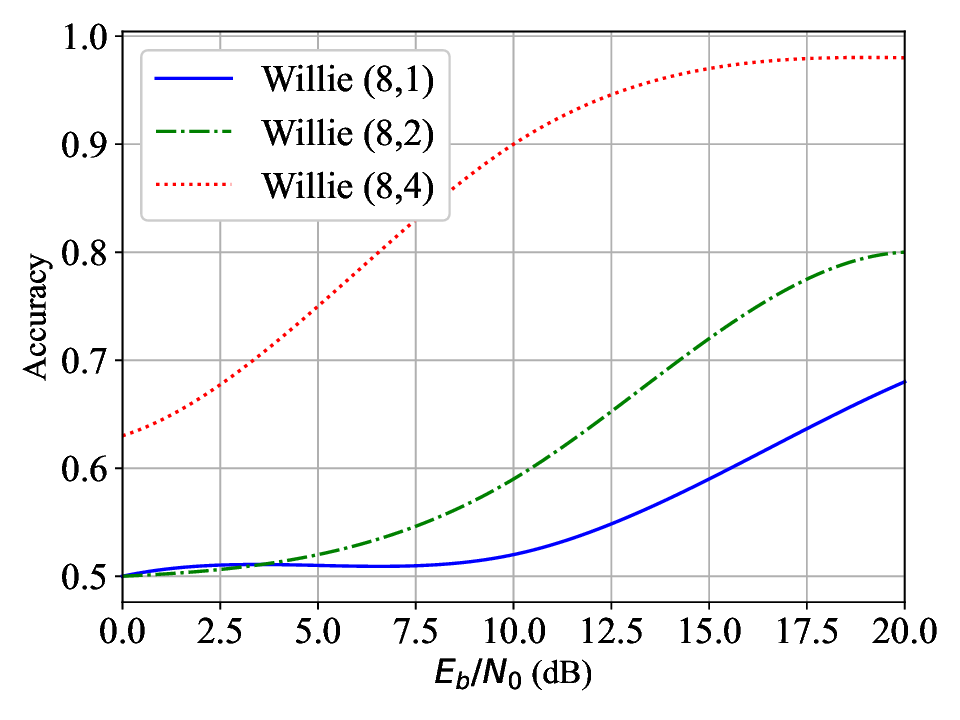}
		\caption{Willie's accuracy}
		\label{fig:rayleigh_resutls_willie}
	\end{subfigure}
	\caption{Single-use model performance over Rayleigh channel for different covert data rates on a range of SNRs.}
	\label{fig:rayleigh_resutls}
\end{minipage}
\begin{minipage}{\textwidth}
	\begin{subfigure}[b]{0.32\textwidth}
		\includegraphics[width=\linewidth]{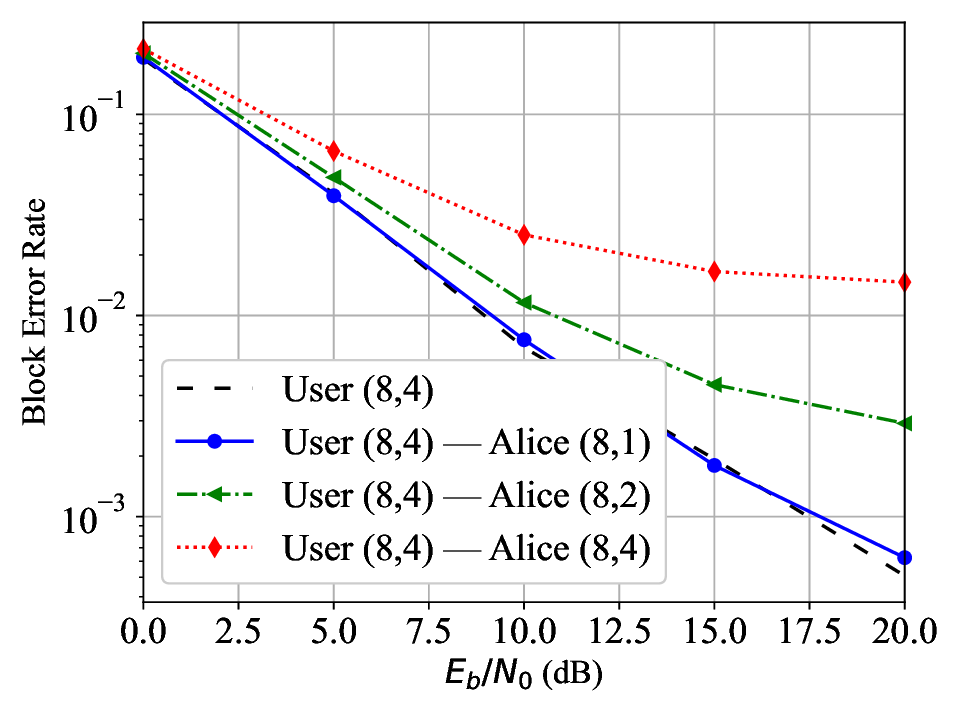}
		\caption{User's BLER}
		\label{fig:rician_resutls_ae}
	\end{subfigure}\hfill
	\begin{subfigure}[b]{0.32\textwidth}
		\includegraphics[width=\linewidth]{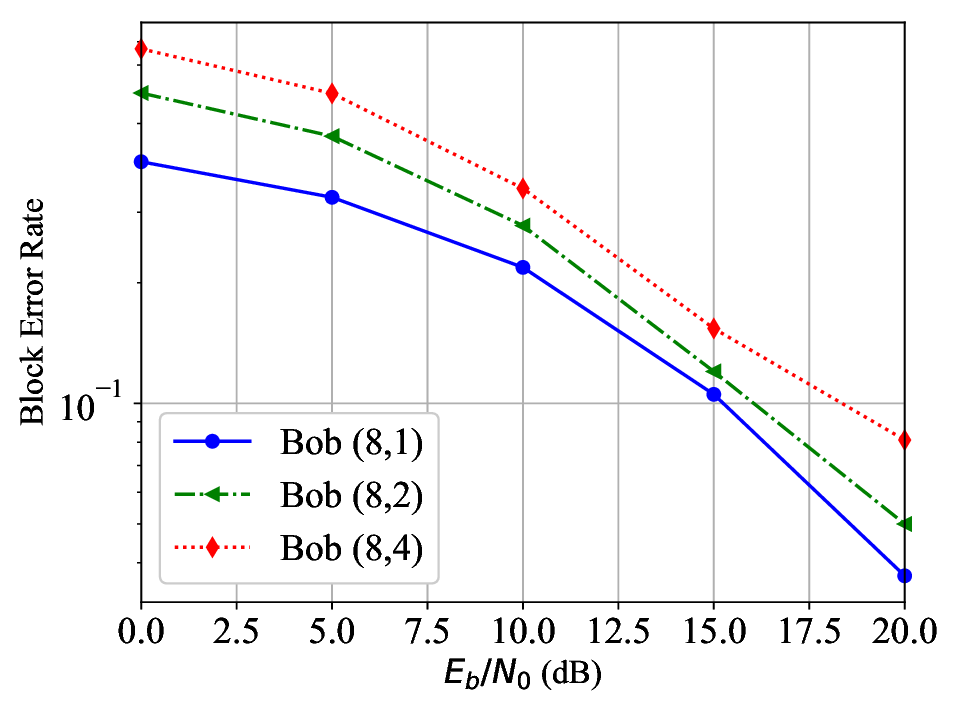}
		\caption{Bob's BLER}
		\label{fig:rician_resutls_bob}	
	\end{subfigure}\hfill
	\begin{subfigure}[b]{0.32\textwidth}
		\includegraphics[width=\linewidth]{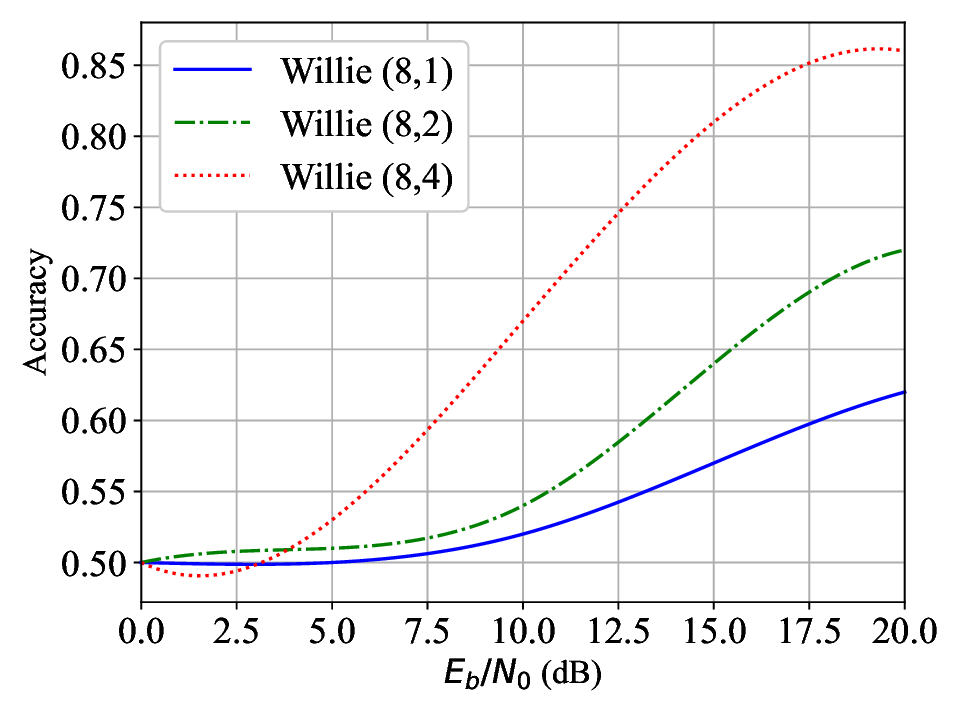}
		\caption{Willie's accuracy}
		\label{fig:rician_resutls_willie}
	\end{subfigure}
	\caption{Single-use model performance over Rician channel for different covert data rates on a range of SNRs.}
	\label{fig:rician_resutls}
\end{minipage}
\end{figure*}

Figs. \ref{fig:awgn_results}, \ref{fig:rayleigh_resutls}, and \ref{fig:rician_resutls} illustrate the performance of our scheme for different covert data rates and how reliable our covert models are at different covert data rates. As we expected, with increasing covert data rates, covert communication becomes more unreliable, its impact on the normal communication increases, and detection becomes easier for Willie. Overall, these plots indicate that sending covert data at high rates makes covert communication unreliable.

The plots also reveal how the communication channel affects the performance of each actor. In the AWGN channel, higher covert rates have a relatively smaller impact on the User's BLER, while in the fading channels, their impact is more significant. On the other hand, increasing the covert rate in the fading channels has less effect on the covert communication accuracy compared to the AWGN channel. For Willie, all channels exhibit a similar trend, where higher covert rates are more susceptible to detection.

Through these experiments, we have concluded that the most reliable covert data rate is achieved by sending 1 bit of data over 8 channel uses. Therefore, we will be using these parameters as the default when evaluating our models in the multi-user scenario.

\textbf{Multi-User Experiments}: 
After evaluating the reliability of our covert models for different covert data rates, we now aim to measure the robustness of our covert scheme against the number of users in the multi-user scenario. To do this, we evaluate our covert models in systems comprising of 2 and 4 users. This will help us understand how adding users, i.e., increasing interference, affects  the performance of our covert models, and whether it has a more significant impact on communication than increasing the covert data rate.

\begin{figure*}
\centering
\begin{minipage}{\textwidth}
	\begin{subfigure}[b]{0.32\textwidth}
		\includegraphics[width=\linewidth]{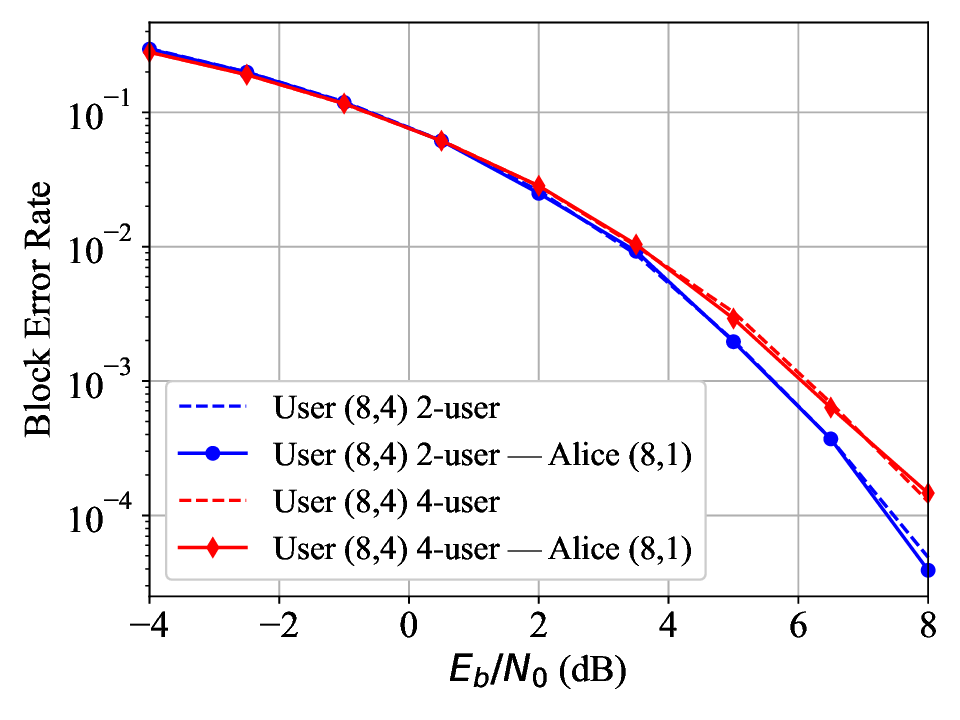}
		\caption{User's BLER}
		\label{fig:multi_awgn_results_ae}
	\end{subfigure}\hfill
	\begin{subfigure}[b]{0.32\textwidth}
		\includegraphics[width=\linewidth]{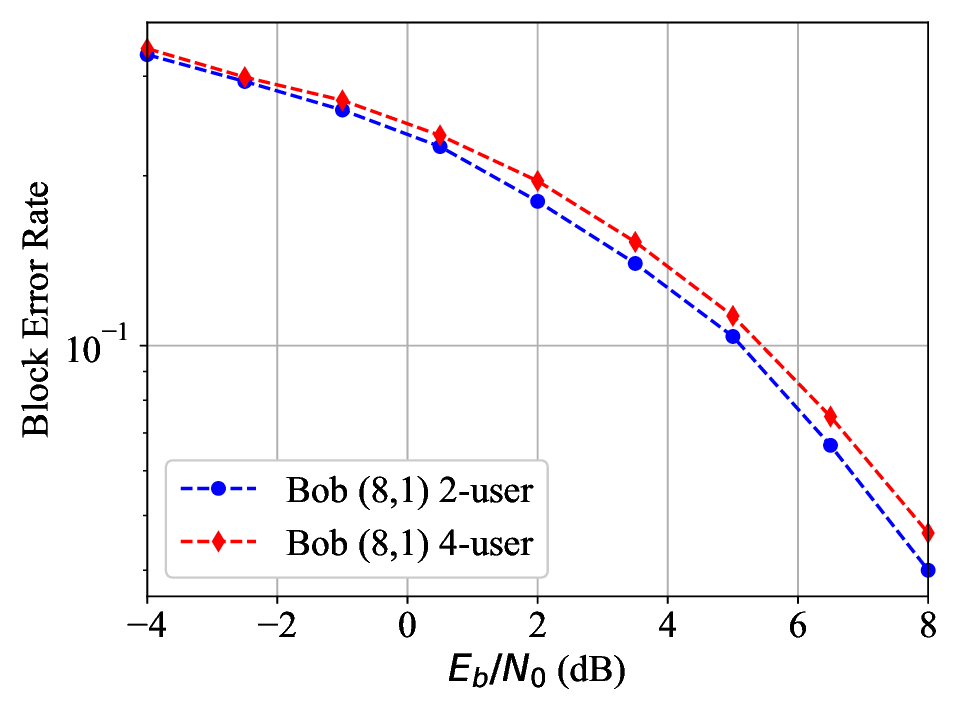}
		\caption{Bob's BLER}	
		\label{fig:multi_awgn_results_bob}
	\end{subfigure}\hfill
	\begin{subfigure}[b]{0.32\textwidth}
		\includegraphics[width=\linewidth]{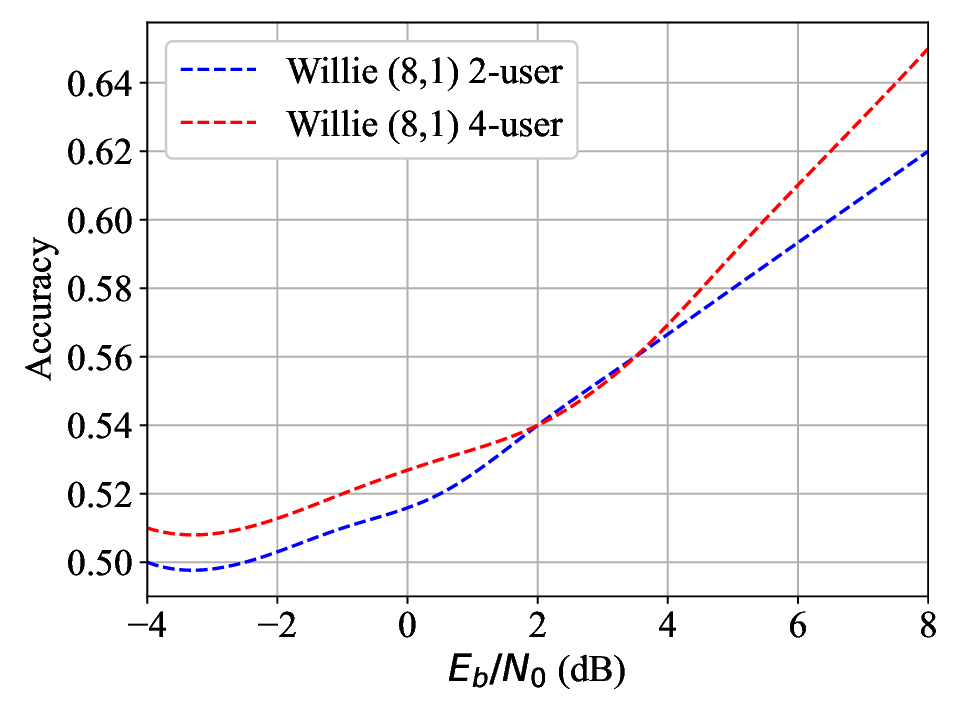}
		\caption{Willie's accuracy}	
		\label{fig:multi_awgn_results_willie}
	\end{subfigure}
	\caption{Multi-user model performance over AWGN channel for systems with different numbers of users.}
	\label{fig:multi_awgn_results}
\end{minipage}
\begin{minipage}{\textwidth}
	\begin{subfigure}[b]{0.32\textwidth}
		\includegraphics[width=\linewidth]{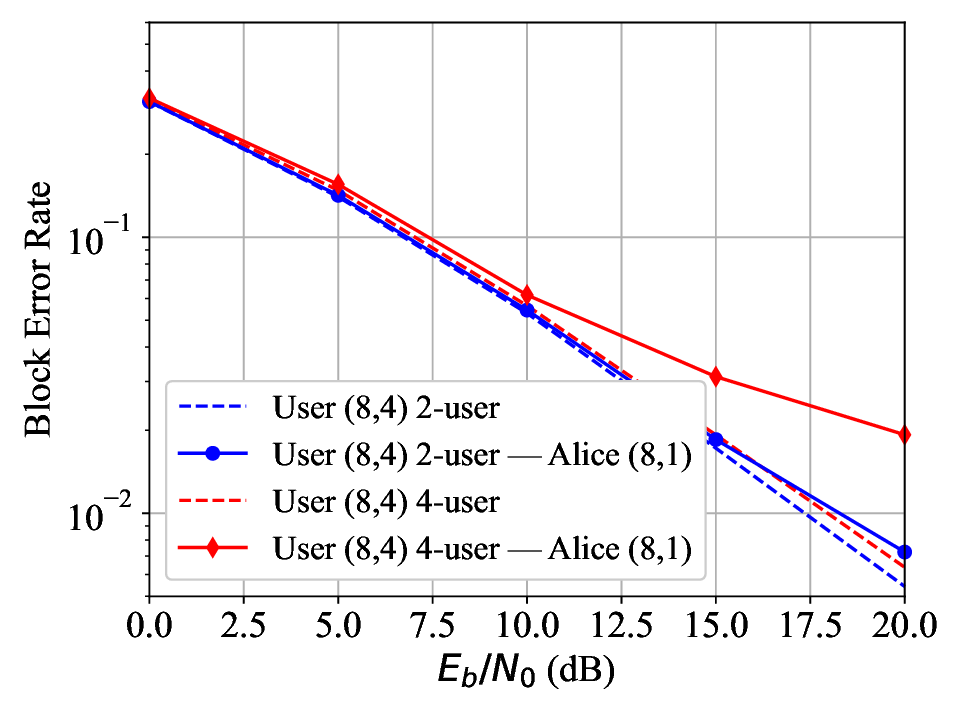}
		\caption{User's BLER}
		\label{fig:multi_rayleigh_results_ae}
	\end{subfigure}\hfill
	\begin{subfigure}[b]{0.32\textwidth}
		\includegraphics[width=\linewidth]{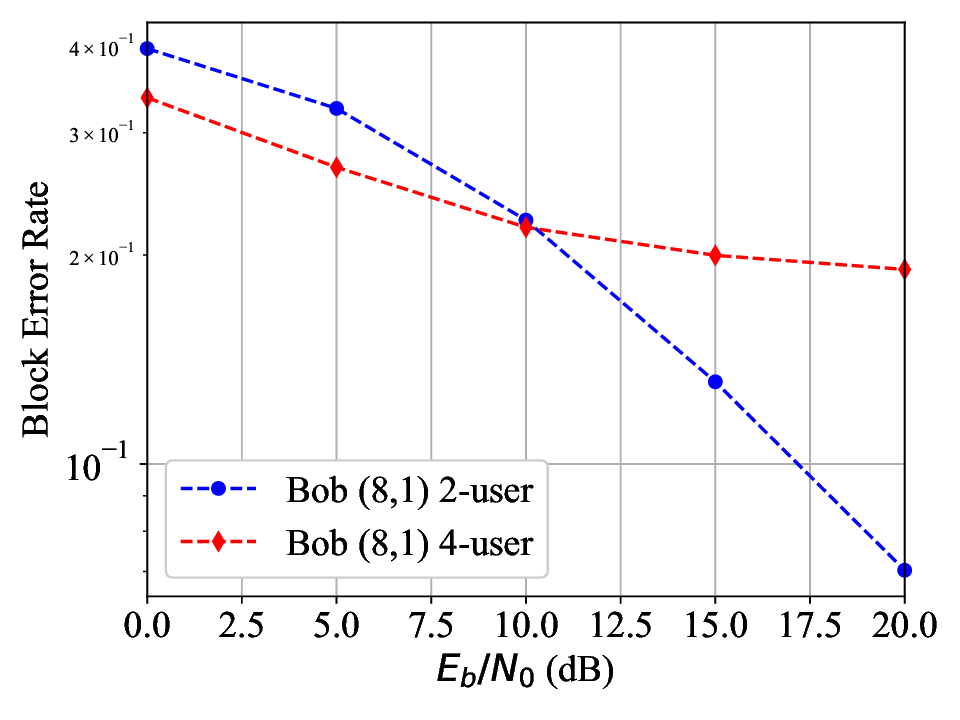}
		\caption{Bob's BLER}
		\label{fig:multi_rayleigh_results_bob}	
	\end{subfigure}\hfill
	\begin{subfigure}[b]{0.32\textwidth}
		\includegraphics[width=\linewidth]{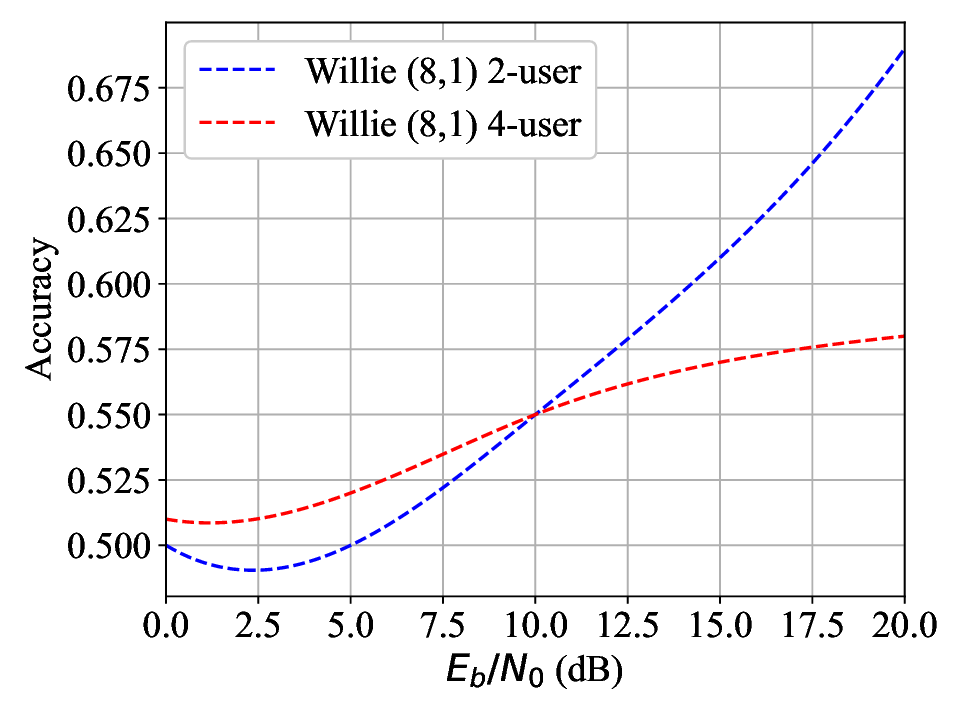}
		\caption{Willie's accuracy}
		\label{fig:multi_rayleigh_results_willie}
	\end{subfigure}
	\caption{Multi-user model performance over Rayleigh channel for systems with different number of users.}
	\label{fig:multi_rayleigh_results}
\end{minipage}
\begin{minipage}{\textwidth}
	\centering
	\begin{subfigure}[b]{0.32\textwidth}
		\includegraphics[width=\linewidth]{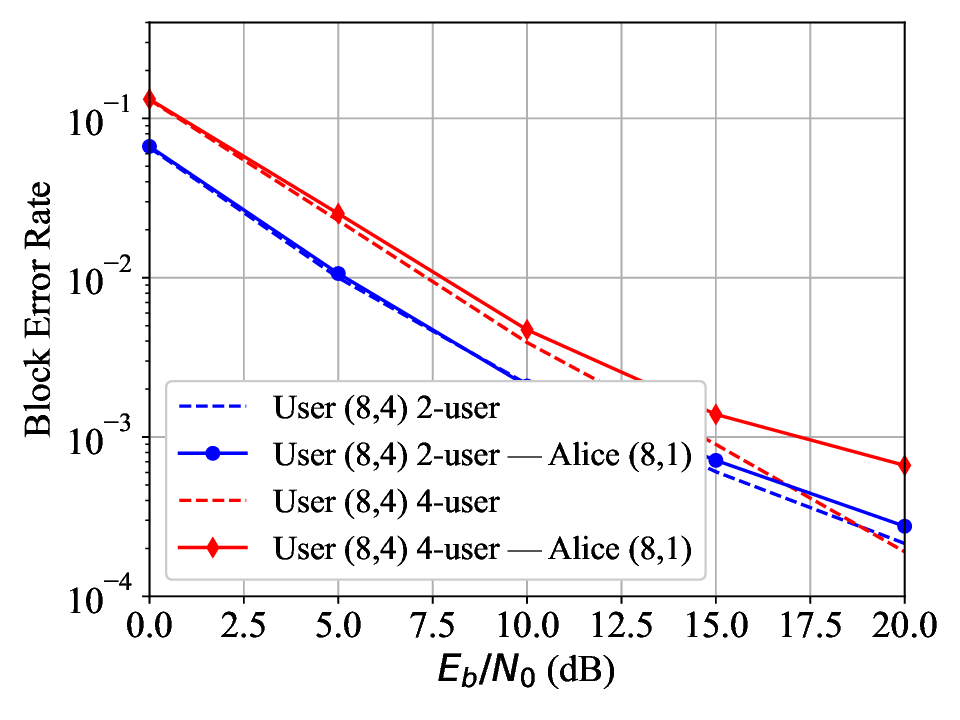}
		\caption{User's BLER}
		\label{fig:multi_rician_results_ae}
	\end{subfigure}\hfill
	\begin{subfigure}[b]{0.32\textwidth}
		\includegraphics[width=\linewidth]{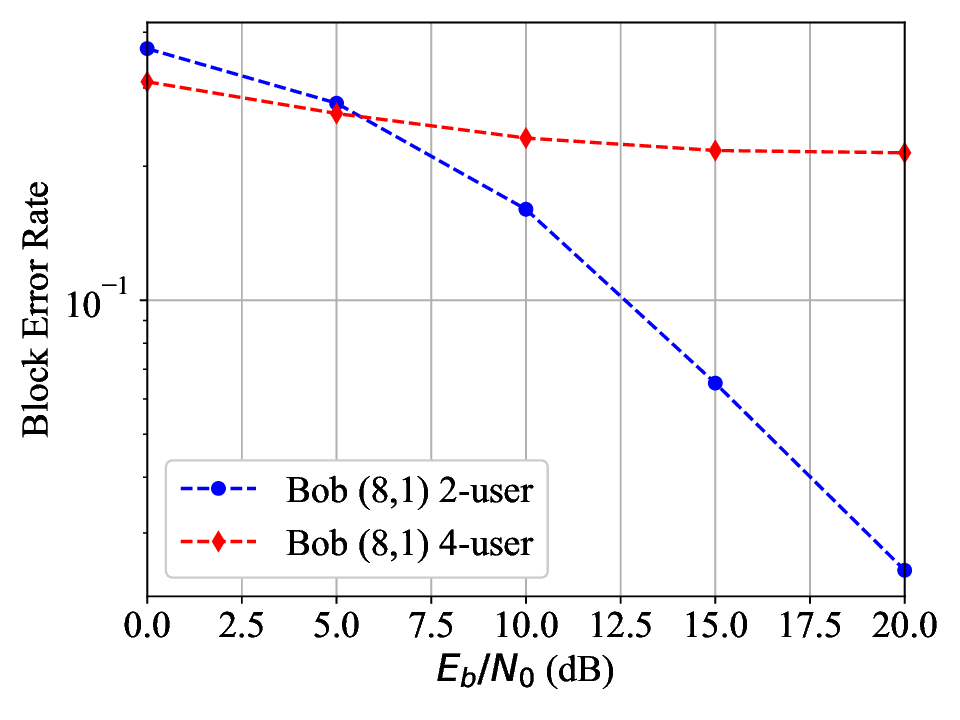}
		\caption{Bob's BLER}
		\label{fig:multi_rician_results_bob}	
	\end{subfigure}\hfill
	\begin{subfigure}[b]{0.32\textwidth}
		\includegraphics[width=\linewidth]{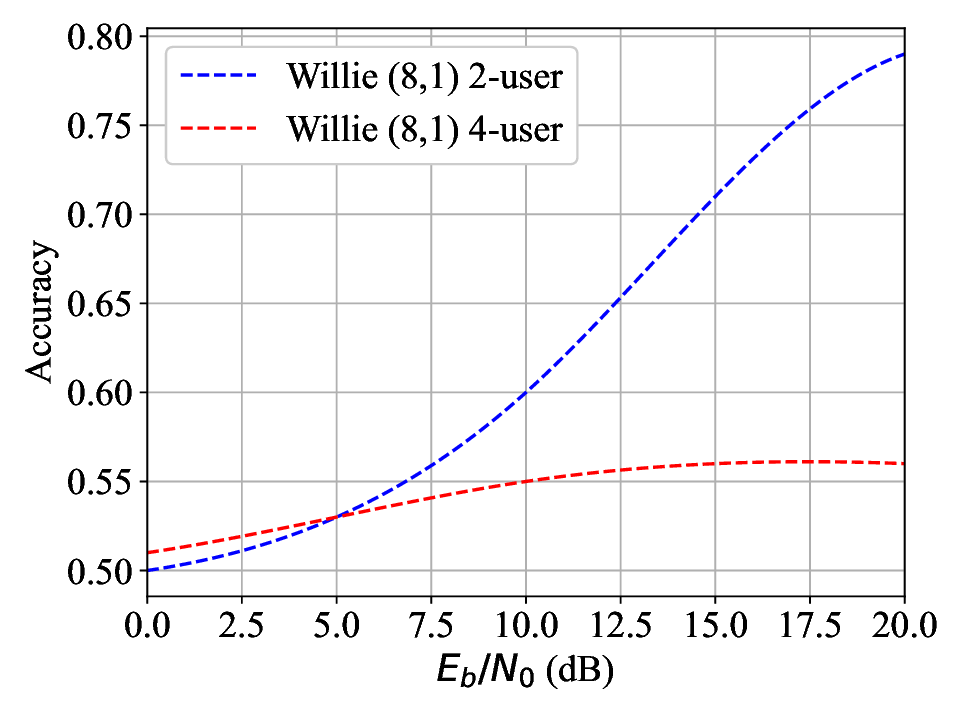}
		\caption{Willie's accuracy}
		\label{fig:multi_rician_results_willie}
	\end{subfigure}
	\caption{Multi-user model performance over Rician channel for systems with different number of users.}
	\label{fig:multi_rician_results}
\end{minipage}
\end{figure*}

Figs. \ref{fig:multi_awgn_results}, \ref{fig:multi_rayleigh_results}, and \ref{fig:multi_rician_results} present our results for 2-user and 4-user systems, demonstrating how the number of users in the system affects our model's performance. For the AWGN channel, we observe that adding more users does not change the model's overall performance. Furthermore, as the number of users increases, there is almost no impact on the normal receivers from the covert transmissions, and Bob and Willie's performances remain almost the same.

However, for the Rayleigh and Rician channels, a degree of freedom effect can be noticed, where increasing number of users makes it more challenging for the covert users to avoid interfering with the ongoing normal communication. With more users, the impact of covert communication on normal users worsens. Unlike in the AWGN channel, adding more users significantly affects Bob's performance in these cases, making covert communication largely ineffective. Looking at Figs. \ref{fig:multi_rayleigh_results} and \ref{fig:multi_rician_results}, we can observe a distinct cross-over pattern for the fading channels. Specifically, Figs. \ref{fig:multi_rayleigh_results_ae} and \ref{fig:multi_rician_results_ae} reveal that there is a certain SNR at which the covert communication in the 4-user systems begins to have a greater impact on normal communication compared to the 2-user systems. These SNRs are 10dB and 5db in the Rayleigh and Rician channels, respectively. These points indicate that covert users can no longer communicate reliably without interfering with normal users. This is evident in Figs. \ref{fig:multi_rayleigh_results_bob} and \ref{fig:multi_rician_results_bob}, where Bob's BLER starts deteriorating at the same SNR values and eventually levels off, deviating from the performance of the 2-user system. Likewise, we can see the same pattern in Willie's detection accuracy. Since covert communication has no specific pattern from these points further, Willie is unable to detect it accurately and thereby his detection accuracy remains constant.

\begin{figure*}[t]
	\center
	\begin{subfigure}{0.325\linewidth}
		\begin{subfigure}{0.48\textwidth}
			\includegraphics[width=\linewidth]{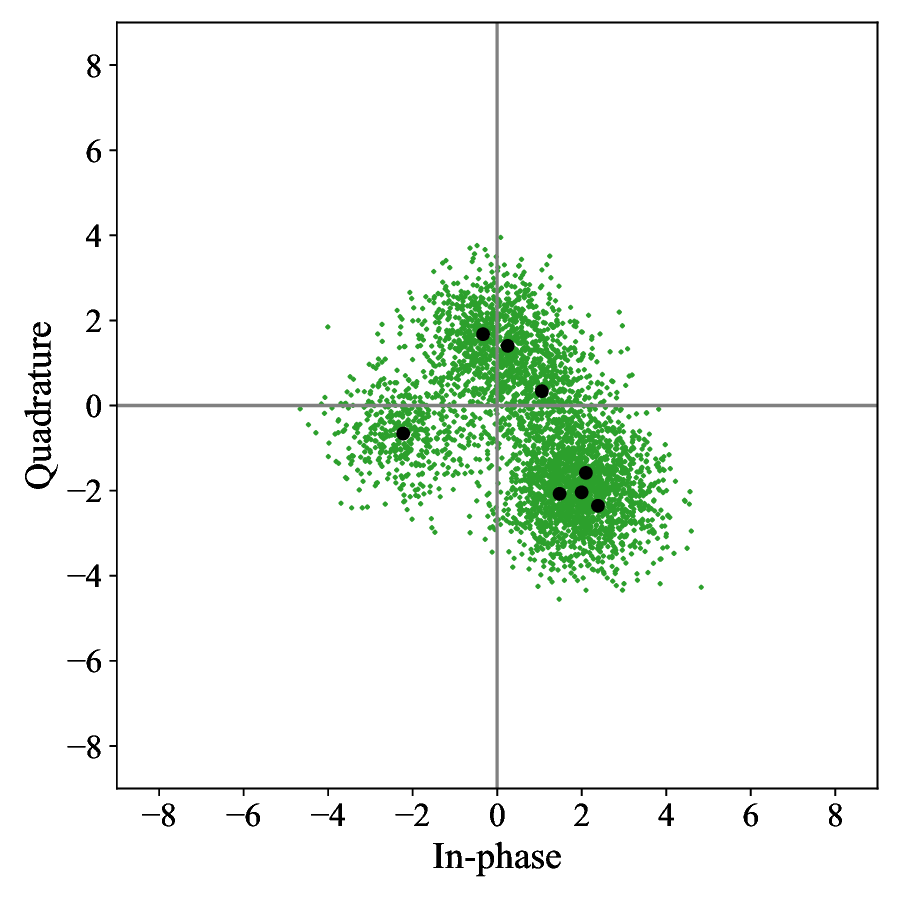}
		\end{subfigure}
		\hfill
		\begin{subfigure}{0.48\textwidth}
			\includegraphics[width=\linewidth]{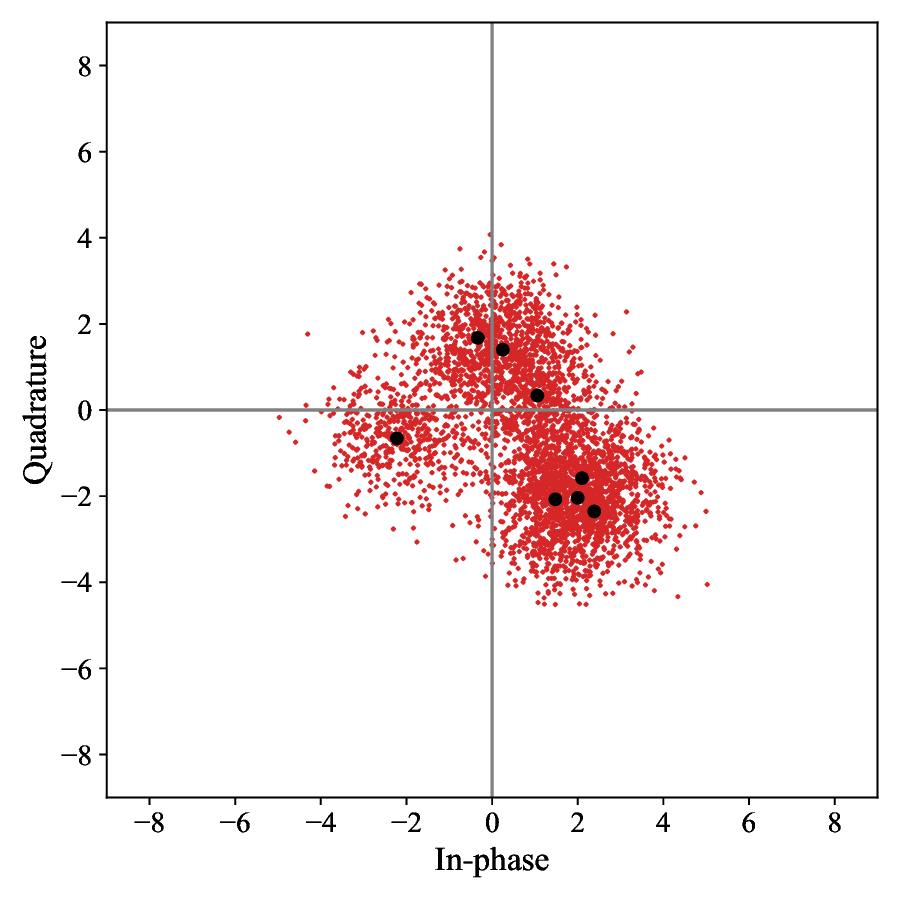}
		\end{subfigure}
		\caption{AWGN channel}
		\label{fig:awgn_constellation}
	\end{subfigure}
	\begin{subfigure}{0.325\linewidth}
		\begin{subfigure}{0.48\textwidth}
			\includegraphics[width=\linewidth]{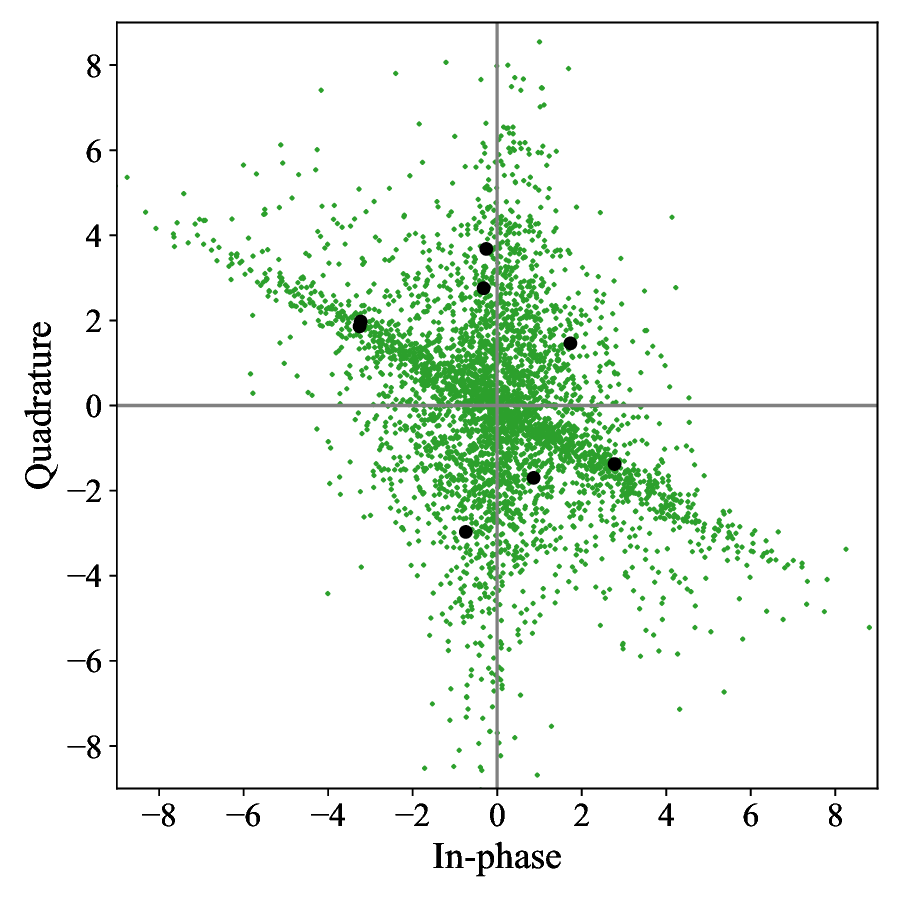}
		\end{subfigure}
		\hfill
		\begin{subfigure}{0.48\textwidth}
			\includegraphics[width=\linewidth]{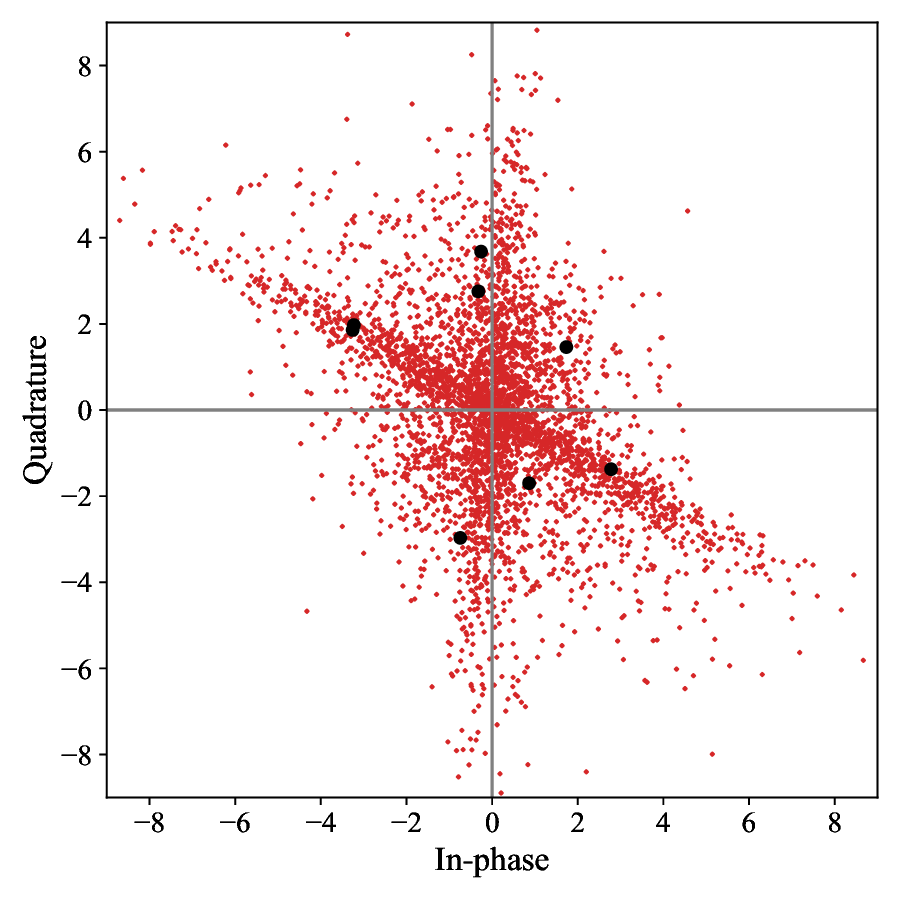}
		\end{subfigure}
		\caption{Rayleigh fading channel}
		\label{fig:rayleigh_constellation}
	\end{subfigure}
	\begin{subfigure}{0.325\linewidth}
		\begin{subfigure}{0.48\textwidth}
			\includegraphics[width=\linewidth]{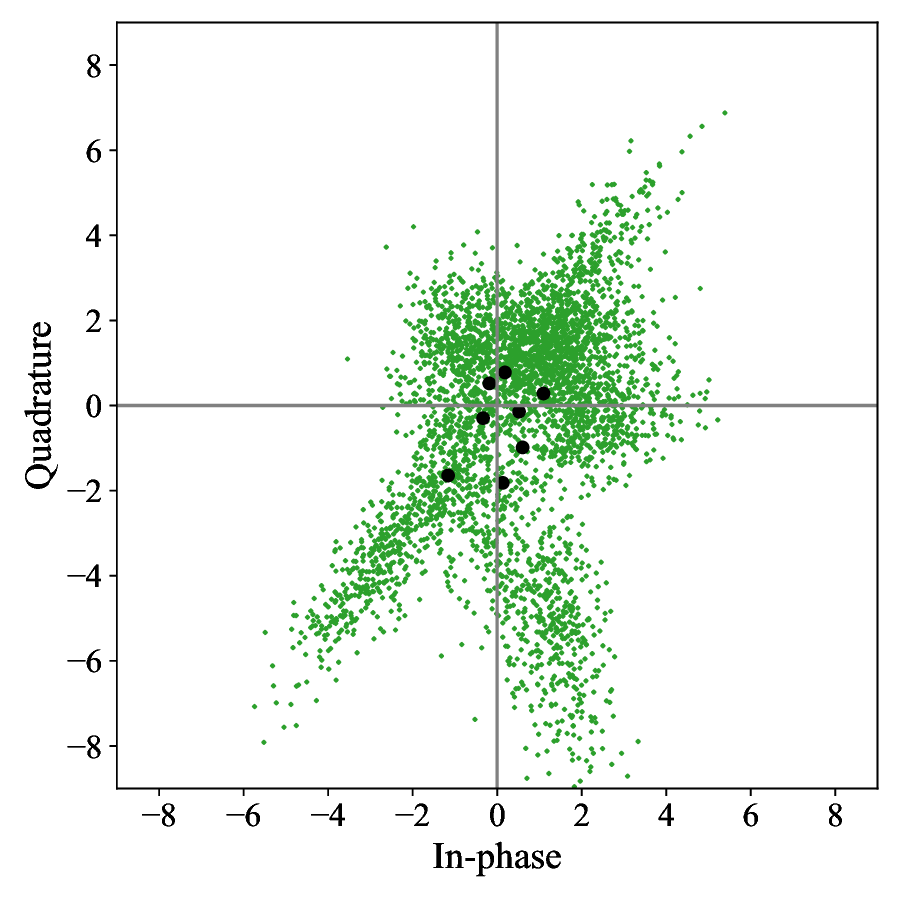}
		\end{subfigure}
		\hfill
		\begin{subfigure}{0.48\textwidth}
			\includegraphics[width=\linewidth]{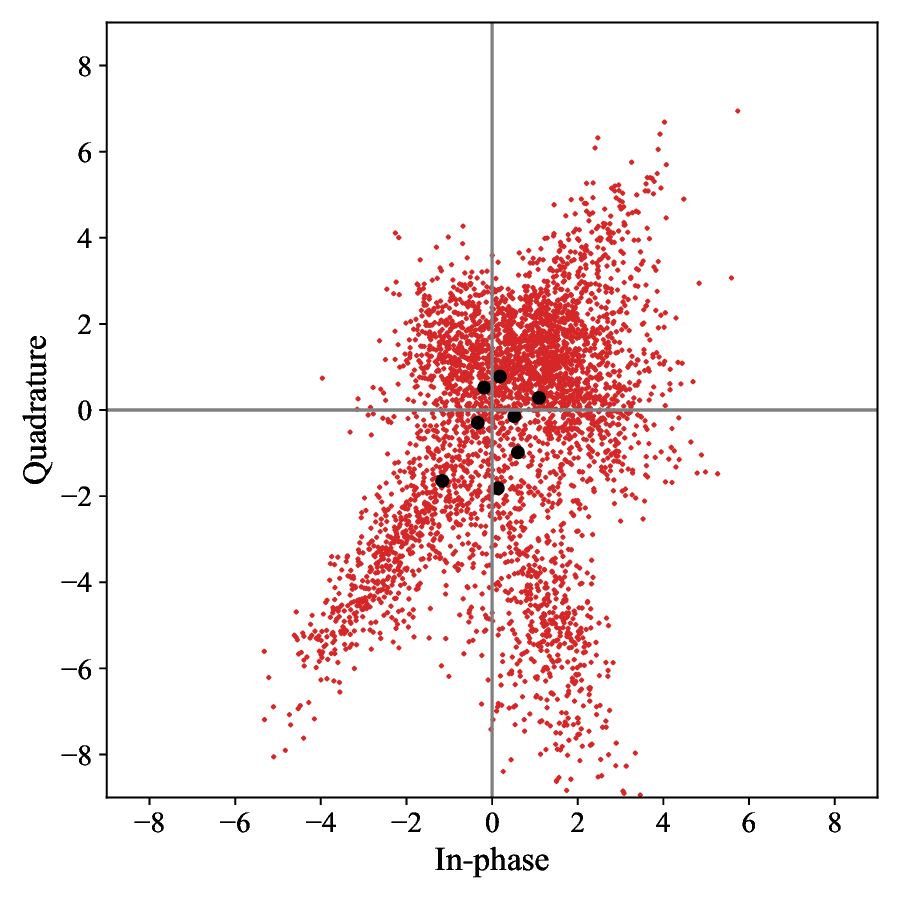}
		\end{subfigure}
		\caption{Rician fading channel}
		\label{fig:rician_constellation}
	\end{subfigure}
	\caption{Comparing AWGN, Rayleigh and Rician fading channels constellation clouds of a sample signal. The green clouds show the constellation before the covert communication and the red clouds show it after.}
	\label{fig:constellation}
\end{figure*}

\textbf{Undetectability}: Willie's detection accuracy can be found in Figs. \ref{fig:awgn_resutls_willie}, \ref{fig:rayleigh_resutls_willie}, and \ref{fig:rician_resutls_willie} for the single-user case, and in Figs. \ref{fig:multi_awgn_results_willie}, \ref{fig:multi_rayleigh_results_willie}, \ref{fig:multi_rician_results_willie} for the multi-user case. His detection performance is evaluated over a range of SNR values for detecting signals as covert and normal. In the single-user experiments, we observe that as the covert data rate increases, the covert communication becomes more easily detectable. In the multi-user case, we cannot directly compare Willie's accuracy for different numbers of users because covert users are unable to establish their covert communication in the fading channels in systems with 4 users. However, the results from the AWGN channel indicate that Willie's accuracy remains roughly the same as we increase the number of users.

\textbf{Constellation Diagrams}: Fig. \ref{fig:constellation} compares the constellation clouds of covert and normal signals for AWGN, Rayleigh, and Rician fading channels in the single-user system. The encoder's output symbols are represented as black circle points on the constellation diagrams. The red cloud represents the scattering of covert signals after passing through the channel, while the green cloud represents this for normal signals. Each chart has 8 black points corresponding to the 8 channel uses. To maintain consistency in Willie's accuracy and Bob's error rate across channel models, we set SNR values to 6dB (AWGN), 15dB (Rayleigh fading), and 16dB (Rician fading). This ensured comparable detection probabilities and a covert communication BLER below \(10^{-1}\). This area of operation provided Alice and Bob relative reliability in their covert communication while maintaining their covertness.
Looking at these figures, the signal constellation diagrams before and after applying our covert model are very similar, showing that to a first-order, Alice has learned to cloak the covert signals into the channel's noise distribution.

	\section{Conclusion}
\label{s:conc}
In this paper, we present a novel deep learning-based covert communication approach that embeds a secret message into a covert signal without the need for handcrafted features. By employing the generative adversarial training framework, we significantly reduce the detection probability of the produced covert signals. Additionally, our proposed training procedure enables us to adjust the trade-off between covert communication reliability and probability of detection by incorporating regularizers in the model's training loss function, regardless of the channel conditions or user messages. Our results show that our covert model is channel-agnostic and insensitive to cover signals. We evaluate the performance of our model across three channel models, namely AWGN, Rayleigh, and Rician fading, and for various covert rates and the number of system users. Furthermore, we investigate the impact of our added covert signals on the ongoing normal system and demonstrate minimal disruption caused by our covert scheme.

Future research could explore the applicability of our covert scheme in blackbox systems, where covert users have no access to the normal users' communication networks. This would involve training our scheme against a substitute normal receiver and evaluating its effectiveness in such scenarios.

	\bibliographystyle{IEEEtran}
	\bibliography{bib/main}


%

\appendix[Model Parameters]
%
%
The following tables summarize the layer configuration and output sizes of various DNN models used in the evaluations.\\

\begin{table*}
		\centering
		\caption{Autoencoder's detailed network architecture in the single-user and multi-user case.}
		\begin{adjustbox}{width=\textwidth}
		\begin{tabular}{c|c|c|c|c|c|}
			\cline{2-6}
			& \textbf{UserTX Encoder} & \textbf{UserRX Parameter Estimation} & \textbf{UserRX Decoder} & \textbf{BaseRX  Pre-Decoder} & \textbf{BaseRX Decoders} \\ \hline
			\multicolumn{1}{|c|}{input size} & 16 & 2 $\times$ 16 & 2 $\times$ 8 & $n_{tx} \times$ 2 $\times$ 8 & $n_{tx} \times$ 4 $\times$ 8 \\ \hline
			\multicolumn{1}{|c|}{dense layers sizes} & 2 $\times$ 8, 2 $\times$ 8 & 2 $\times$ 16, 2 $\times$ 32, 2 $\times$ 8 & 2 $\times$ 8, 2 $\times$ 8 & \begin{tabular}[c]{@{}c@{}}$n_{tx} \times$ 2 $\times$ 8, $n_{tx} \times$ 4 $\times$ 8\end{tabular} & $n_{tx} \times$ 2 $\times$ 8 \\ \hline
			\multicolumn{1}{|c|}{dense layers activations} & 2 $\times$ ELU & ELU, 2 $\times$ Tanh & 2 $\times$ Tanh, Softmax & 3 $\times$ Tanh & Tanh, Softmax \\ \hline
			\multicolumn{1}{|c|}{conv filters} & 1, 8, 8, 8 & - & 1, 8, 8, 8 & 1, 8, 8, 8 & - \\ \hline
			\multicolumn{1}{|c|}{conv kernel sizes} & 2, 4, 2, 2 & - & 2, 4, 2, 2 & 2, 4, 2, 2 & - \\ \hline
			\multicolumn{1}{|c|}{conv strides} & 1, 2, 1, 1 & - & 1, 2, 1, 1 & 1, 2, 1, 1 & - \\ \hline
			\multicolumn{1}{|c|}{conv activations} & 4 $\times$ Tanh & - & 4 $\times$ Tanh & 4 $\times$ Tanh & - \\ \hline
			\multicolumn{1}{|c|}{output size} & 2 $\times$ 8 & 2 $\times$ 1 & 16 & $n_{tx} \times$ 4 $\times$ 8 & 16 \\ \hline
		\end{tabular}
		\end{adjustbox}
	\label{table:autoencoder_structure}
\end{table*}
\begin{table*}
		\centering
		\caption{Alice, Bob, and Willie's detailed network architecture in the single-user and multi-user case.}
		\begin{adjustbox}{width=\textwidth}
		\begin{tabular}{c|c@{}|c@{}|c@{}|c@{}|c|}
			\cline{2-6}
			\multicolumn{1}{l|}{} & \textbf{Alice (Single-User)} & \textbf{Alice (Multi-User) AWGN} & \textbf{Alice (Multi-User) Rayleigh/Rician} & \textbf{Bob} & \textbf{Willie} \\ \hline
			\multicolumn{1}{|c|}{input size} & 16 + $2^{k}$ & 16  + $2^{k}$ & 16  + $2^{k+1}$ + ($n_{tx} \times n_{tx} \times 2$) & 16 & 16 \\ \hline
			\multicolumn{1}{|c|}{dense layers sizes} & \begin{tabular}[c]{@{}c@{}}32  + $2^{k+1}$, \\ 32 + $2^{k+1}$, \\ 8 $\times 2^k$\end{tabular} & \begin{tabular}[c]{@{}c@{}}32  + $2^{k+1}$, \\ 32 + $2^{k+1}$, \\ 8 $\times 2^k$\end{tabular} & \begin{tabular}[c]{@{}c@{}}32  + $2^{k+1}$ + ($n_{tx} \times n_{tx} \times 2$),\\ 32 + $2^{k+1}$, \\ 32 + $2^{k+1}$, \\ 8 $\times 2^k$\end{tabular} & 2 $\times$ 8, 16 & 2 $\times$ 8, 2 $\times$ 8 \\ \hline
			\multicolumn{1}{|c|}{dense layers activations} & 3 $\times$ ReLU, Tanh & 3 $\times$ ReLU, Tanh & 4 $\times$ ReLU, Tanh & 2 $\times$ Tanh, Softmax & 2 $\times$ Tanh, Sigmoid \\ \hline
			\multicolumn{1}{|c|}{conv filters} & - & - & - & 1, 8, 8, 8, 8 & 1, 8, 8, 8, 8 \\ \hline
			\multicolumn{1}{|c|}{conv kernel sizes} & - & - & - & 1, 2, 4, 2, 2 & 1, 2, 4, 2, 2 \\ \hline
			\multicolumn{1}{|c|}{conv strides} & - & - & - & 1, 1, 2, 1, 1 & 1, 1, 2, 1, 1 \\ \hline
			\multicolumn{1}{|c|}{conv activations} & - & - & - &  5 $\times$ LeakyReLU & 5 $\times$ LeakyReLU \\ \hline
			\multicolumn{1}{|c|}{output size} & 2 $\times$ 8 & 2 $\times$ 8 & 2 $\times$ 8 & $2^{k}$ & 1 \\ \hline
		\end{tabular}
		\end{adjustbox}
	\label{table:covert_models_structure}
\end{table*}

\end{document}